\documentclass[aps,pre,superscriptaddress,reprint]{revtex4-2}
\usepackage{graphicx,color,hyperref} 
\usepackage{verbatim} 
\usepackage{epsfig,amsmath,amssymb,calc,verbatim}
\usepackage{bm}
\usepackage{subcaption}
\usepackage{ragged2e}
\usepackage{physics}

\newcommand{\sr}{f_r} 
\newcommand{\nr}{f_N} 
\newcommand{\calO}{{\cal O}} %
\begin{document}

\preprint{ATP'S/123-QED}

\title{Hyperuniformity near jamming transition over a wide range of bidispersity}
\author{Duc T. Dam}
\email{dam@r.phys.nagoya-u.ac.jp}
\affiliation{Department of Physics, Nagoya University, Nagoya 464-8602, Japan}
\author{Takeshi Kawasaki}
\affiliation{Department of Physics, Nagoya University, Nagoya 464-8602, Japan}
\affiliation{D3 Center, The University of Osaka, Toyonaka, Osaka 560-0043, Japan}
\affiliation{Department of Physics, The University of Osaka, Toyonaka, Osaka 560-0043, Japan}
\author{Atsushi Ikeda}
\affiliation{Graduate School of Arts and Sciences, The University of Tokyo, Tokyo 153-8902, Japan}
\affiliation{Research Center for Complex Systems Biology, Universal Biology
Institute, The University of Tokyo, Tokyo 153-8902, Japan} 
\author{Kunimasa Miyazaki}
\email{miyazaki@r.phys.nagoya-u.ac.jp}
\affiliation{Department of Physics, Nagoya University, Nagoya 464-8602, Japan}

\date{\today}

\begin{abstract}
We numerically investigate hyperuniformity in two-dimensional 
jammed packings of frictionless bidisperse particles. 
Hyperuniformity is characterized by the suppression of density fluctuations at
 large length scales, and the structure factor vanishes at small wavenumbers as $S(q) \propto q^{\alpha}$, where $\alpha > 0$. 
Jammed packings are known to exhibit hyperuniformity over a wide wavenumber
 window, down to $q^{\ast}\sigma \approx 0.2$,  
where $\sigma$ is the particle diameter. 
In two dimensions, we find that the exponent $\alpha$ is approximately
 $0.6\text{--}0.7$, in contrast to the reported value $\alpha = 1$ in 
 three-dimensional systems. 
We then extend the analysis to a wide range of particle size ratios, from 
the monodisperse limit to highly disparate mixtures.
To reduce unwanted spatial heterogeneities associated with polycrystalline
 domains at very large or very small size disparities,
we employ a recently proposed method by Rissone \textit{et al.}
 \href{https://link.aps.org/doi/10.1103/PhysRevLett.127.038001}{[Phys. Rev. Lett. {\bf
 127}, 038001 (2021)]}.
This method enables a more precise determination of $\alpha$.
We find that $\alpha$ remains nearly constant over the entire range of size
 ratios, except in the monodisperse case, where crystallization occurs. 
\end{abstract}

\maketitle

\section{\label{sec:intro} Introduction}

Hyperuniform systems are a distinct state of matter in which density
fluctuations are anomalously suppressed at large length scales
\cite{torquato2003local, torquato2018hyperuniform}. For equilibrium systems in
$d$ dimensions, the particle-number fluctuations
${\ev{\delta N^2}=\ev{N^2}-\ev{N}^2}$ within a subsystem of size $R$ are
extensive, {i.e.,} 
${\ev{\delta N^2} \propto \ev{N} \propto R^d}$.  
Equivalently, in Fourier space, the static structure factor 
${S(q) =\frac{1}{N}\ev{|\delta\rho(\mathbf{q})|^2}}$, where 
$\delta\rho(\mathbf{q})$ is the density fluctuation at wavenumber $\mathbf{q}$,
approaches a constant in the small-wavenumber limit $q \equiv |\mathbf{q}| \to 0$.  
In contrast, for hyperuniform systems, $\ev{\delta N^2}$ 
becomes sub-extensive and scales in the large-$R$ limit as $R^{\beta}$ with $\beta<d$, while $S(q)$ vanishes in the small-$q$ limit as
\begin{equation}
	\label{eq:HUasymtotic_S}
	S(q) \propto q^{\alpha}  \quad \text{for} \quad q \to 0, 
\end{equation}
where $\alpha > 0$ is the \textit{hyperuniformity exponent} that characterizes the degree of hyperuniformity. The exponent $\beta$ is related to $\alpha$ by $\beta=d-\alpha$ if $0<\alpha<1$.

Over the past few decades, hyperuniformity has been identified in a variety of nonequilibrium systems, including fluctuations in the early universe \cite{gabrielli2002glass}, avian photoreceptor patterns \cite{jiao2014avian}, receptor organization in the immune system \cite{mayer2015well}, and glasses \cite{zhang2016perfect}. The packing of frictionless spheres near the jamming transition is among the earliest known examples of hyperuniform systems~\cite{torquato2003local, donev2005pair, zachary2011hyperuniform, zachary2011hyperuniformity, atkinson2016critical}.

The jamming transition occurs when athermal soft particles cease flowing and the
system acquires rigidity, forming an amorphous solid as the packing fraction
$\varphi$ increases beyond a critical value $\varphi_J$~\cite{liu1998jamming,
o2003jamming, van2009jamming}. The jamming transition of frictionless particles
is characterized by several critical behaviors \cite{o2003jamming,
van2009jamming}. 
Examples include the algebraic dependence of various
quantities, such as the contact number $\delta z$ and shear modulus $G$, on the
distance to jamming $\delta\varphi = \varphi - \varphi_J$, as well as the
emergence of diverging collective length scales and universal low-frequency
vibrational modes. Although these critical behaviors have been well explained by
geometric variational arguments~\cite{Wyart2005} and mean-field
theory~\cite{charbonneau2014fractal}, 
the hyperuniform behavior of jammed systems
remains outside the scope of these arguments, and the connection between the
jamming transition and hyperuniformity is poorly understood.

Hyperuniformity in jammed packings was first discussed in
Ref.~\cite{torquato2003local}, which conjectured that infinitely large strictly
jammed saturated packings in $d$ dimensions are hyperuniform. 
Numerical studies of maximally random jammed (MRJ) packings, as defined in
Ref.~\cite{torquato2000random}, first demonstrated such behavior,
showing that the structure factor scales linearly as $S(q) \propto q$ ({i.e.,}
$\alpha=1$) \cite{donev2005unexpected}. 
Since then, numerous studies have shown
evidence of hyperuniform behavior in jammed packings generated via different
protocols and extended to binary mixtures by introducing several variants 
of $S(q)$~\cite{donev2005pair,berthier2011suppressed,wu2015search,dreyfus2015diagnosing,ozawa2017exploring,chieco2018spectrum}.  

A closer inspection of the small-wavenumber behavior of $S(q)$ reveals that the
hyperuniform scaling with $\alpha=1$ persists only down to a finite value of 
$q^{\ast}$, which is approximately $0.2\sigma^{-1}$ ($\sigma$
is the particle diameter) for both $d=2$ and $3$~\cite{wu2015search,ikeda2017large}. 
$S(q)$ at $q < q^{\ast}$ appears to depend sensitively on the
preparation protocols;  
for the packings prepared using standard protocols, $S(q)$ either saturates or even
increases below
$q^{\ast}$~\cite{wu2015search,ikeda2015thermal,ikeda2017large,ozawa2017exploring,berthier2011suppressed,xu2010effects},
while if the jamming transition is approached dynamically 
using a random organization model, the hyperuniformity is reproduced but with 
the distinct exponent $\alpha=0.25$ for $d=3$ and $0.45$ for
$d=2$~\cite{Henkel2008book,wilken2021random,wilken2023dynamical}.
Though the fate of the hyperuniformity below $q^{\ast}$ is a controversial issue, 
the power-law scaling $S(q) \propto q$ at $q > q^{\ast}$ for $d=3$ is robustly observed and
has been unquestioned so far. 
 
The goal of the present study is to revisit the hyperuniformity 
${S(q) \propto q^{\alpha}}$ at $q > q^{\ast}$ in two-dimensional systems. 
We consider a binary mixture of large and small particles of size 
$\sigma_L$ and $\sigma_S$, respectively, to avoid crystallization. 
Recently, we investigated density hyperuniformity in this intermediate-$q$ regime
in two-dimensional binary mixtures near $\varphi_J$ and reported a smaller
hyperuniformity exponent, $\alpha \approx 0.6\text{--}0.7$~\cite{matsuyama2021geometrical}. 
This value has proven robust and insensitive to different annealing protocols used to prepare the packings. 
Surprisingly, this deviation from the exponent $\alpha=1$ reported for the three-dimensional system has received little attention in the literature.  
In this study, we first re-examine the hyperuniformity exponent for
two-dimensional packings for an equimolar bidisperse system 
with a size ratio of 
${\sigma_S: \sigma_L=1{:}1.4}$, 
the most widely used model system in studies of the jamming transition. 
Then we extend the analysis to a broad range of
size ratios, $\sr\equiv \sigma_S/\sigma_L$, from as small as 0.3
to the monodisperse limit of $1$. 
It is known that the particle size ratio significantly affects the
jamming transition density $\varphi_J$ and local structural
features~\cite{koeze2016mapping, saitoh2025jamming,ricouvier2017optimizing}. 
When the size ratio $\sr$ is close to 1, the packing tends to develop crystalline order.
In contrast, when $\sr$ is small, the small particles tend to be embedded in the void spaces
formed by the large particles, which enhances the concentration fluctuations.
Surprisingly, recent studies have shown that the critical scaling properties for
typical observables such as $\delta z$ and $G$ remain
robust and survive for a wide range of the size ratio
$\sr$~\cite{goodrich2014solids,kawasaki2024unified,pan2023review,saitoh2025jamming}. 
Notably, jamming criticality persists near $\varphi_J$ even as the size ratio
approaches unity and the system becomes increasingly crystalline, although  
the range over which critical scaling is observed narrows progressively in the
monodisperse limit, $\sr \to 1^{-}$~\cite{tong2015crystals, ikeda2020jamming}, 
except for a singular point $\sr = 1$ at which the system forms an ideal crystal
with no jamming criticality~\cite{tong2015crystals}. 
It is therefore natural to expect that the hyperuniformity should also persist  
for a range of $\sr$. 

In general, observing hyperuniformity in mixtures is more difficult compared
to monodisperse systems. 
The simple structure factor $S(q)$, defined from the particle density field 
$\rho(\mathbf{q}) =\sum_{j} e^{-i\mathbf{q}\cdot\mathbf{r}_j}$,
does not clearly exhibit hyperuniformity, and several alternative measures, such as the generalized compressibility and the correlations of the local volume fraction, have been proposed~\cite{torquato2018hyperuniform,berthier2011suppressed,wu2015search, zachary2011hyperuniformity}. 
However, when $\sr$ is either too small or too close to unity, even these
variants become ineffective: 
systems with large size disparity show strong compositional fluctuations, 
whereas systems close to the monodisperse limit develop large
polycrystalline domains and become spatially heterogeneous~\cite{ricouvier2017optimizing, maher2023hyperuniformity,dreyfus2015diagnosing}. 
To reduce unwanted local compositional fluctuations at
small $\sr$ and noise arising from grain boundaries as $\sr \to 1^-$, 
we introduce a method inspired by Rissone \textit{et
al.}~\cite{rissone2021long}, originally developed for monodisperse 
three-dimensional systems. 
Their approach exploits the duality between particle centers and
contact points in isostatic jammed networks.  
By shifting the representation from particle centers to contact points, they measured 
a hyperuniformity exponent $\alpha = 1$ in the force and mutual contact-number
fluctuations in real space for the monodisperse systems. 
Their result provides supporting evidence that hyperuniformity of the real
particles at the jamming transition is inherited by the contact points (which we
refer to as quasi-particles hereafter).

In this study, we adopt their method for bidisperse systems in two
dimensions.  
Strictly speaking, this method cannot be applied to systems with
size disparity, as the duality between particle centers and contact points no
longer holds when the size ratio deviates from unity. 
However, we find that this approach significantly improves the
accuracy with which the hyperuniformity exponent $\alpha$ can be determined,
provided that appropriate weighting variables associated with the 
quasi-particles are chosen. 
Using this method, we show that the hyperuniformity scaling persists for 
$q > q^{\ast}$ with the crossover wavenumber $q^{\ast}$ remaining essentially unchanged,
and the hyperuniformity exponent $\alpha$ 
consistently falls within the range $0.6 \lesssim \alpha \lesssim 0.7$
over the entire range of size ratios from the disparate value $\sr=0.3$ up to the
nearly monodisperse limit $\sr \to 1^{-}$, excluding the singular case $\sr=1$. 

The remainder of this paper is organized as follows. In Sec.~\ref{sec:model}, we describe the details of our simulation model and the protocol used to generate the jammed packings. In Sec.~\ref{sec:theory}, we outline the concepts underlying the quasi-particle representation approach introduced by Rissone \textit{et al.}~\cite{rissone2021long} and our extension of this method to bidisperse systems. In Sec.~\ref{sec:ResultsandDiscussions}, we present our numerical results. We conclude this paper in Sec.~\ref{sec:conclusionanddiscussion}.

\section{\label{sec:model}Method and Simulation Protocol}

We study jammed packings of frictionless soft disks using numerical
simulations~\cite{van2009jamming, o2003jamming}. The system is a binary mixture
of small and large particles with diameters $\sigma_S$ and $\sigma_L$,
respectively.  
We fix the number fraction at $\nr = N_S / (N_S + N_L) = 0.5$, where $N_S$ and
$N_L$ denote the numbers of small and large particles.  
Most of the results presented below are based on systems with a total of 
$N =N_S + N_L = 4\,000$ particles. 
We also perform simulations with $N = 10\,000$ to
confirm that our results are not affected by finite-size effects. To explore a
broad range of structural orderings, we vary the particle size ratio $\sr =
\sigma_S / \sigma_L$ from $0.3$ to $1.0$, following an approach similar to that
used in previous studies~\cite{ricouvier2017optimizing, koeze2016mapping}. Here,
$\sr = 1$ corresponds to a monodisperse system. The size ratio $\sr = 1/1.4
\approx 0.71$ is the most commonly used in studies of the jamming
transition~\cite{o2003jamming, wu2015search}.  
The $i$-th and $j$-th particles interact via a harmonic potential
\begin{equation}
\label{potentialform}
U(r_{ij})= 
\frac{\epsilon}{2}\left(1-\frac{r_{ij}}{\sigma_{ij}}\right)^2 H\left( 1-\frac{r_{ij}}{\sigma_{ij}}\right),
\end{equation}
where ${r_{ij}=\left|\mathbf{r}_i-\mathbf{r}_j\right|}$ 
is the interparticle distance,  
${\sigma_{ij} =(\sigma_i + \sigma_j)/2}$ 
is the average diameter of particles $i$ and $j$, 
and $H(x)$ denotes the Heaviside step function.
We use $\sigma_S$ and $\epsilon$ as the units of length and energy, respectively. 

All measurements of structure factors and density fluctuations in this study are
based on jammed configurations slightly above the jamming transition 
($\varphi \gtrsim \varphi_J$),
generated using the protocol described below. 
We first prepare a
random configuration at a low packing fraction, 
$\varphi_{\text{ini}}=0.835$. 
The system is then compressed quasistatically with a packing-fraction increment
$\Delta \varphi$. After each compression step, the system is relaxed 
to a local energy minimum using the FIRE
algorithm~\cite{bitzek2006structural}. 
We consider a configuration to be mechanically stable when the average
force amplitude acting on a particle is less than 
$10^{-14} \epsilon/\sigma_S$. 
This cutoff reflects the numerical
precision limit of double-precision arithmetic and round-off errors. 
A configuration is considered jammed when the average potential energy per
particle becomes nonzero, specifically when 
$N^{-1}\sum_{j>k} U(r_{jk}) >10^{-16}$. 
To obtain a jammed configuration just above the jamming transition
density $\varphi_J$, we start compressing with 
$\Delta \varphi = 2 \times 10^{-3}$. 
Whenever the compression overshoots $\varphi_J$, we return to the
last unjammed state and repeat the compression with $\Delta \varphi$ reduced by
half. This iterative procedure continues until 
$\Delta \varphi < 10^{-6}$,
ensuring that the final jammed configuration satisfies 
$0 < (\varphi-\varphi_J) < 10^{-6}$. 
This protocol has been widely used in the
past~\cite{o2003jamming} and closely resembles experimental procedures in
granular systems~\cite{majmudar2007jamming}. 

\begin{figure}[tb]
\centering
\includegraphics[width=0.5\textwidth]{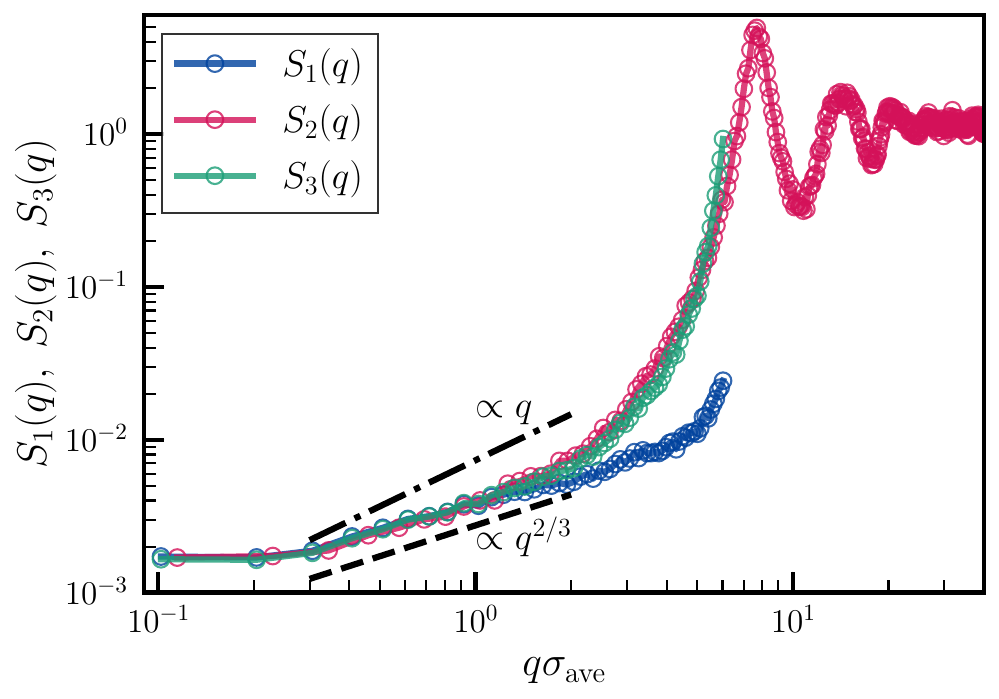}
\caption{\label{fig:threedefinition} 
\justifying 
Fourier-space measures of density fluctuations for bidisperse jammed packings with
$\sr = 0.71$ defined by Eqs.~(\ref{def1}), (\ref{def2}), and (\ref{def3}).
}
\end{figure}

Let us recapitulate the hyperuniformity of the most commonly studied jammed system with a size ratio of $\sr = 1/1.4 \approx 0.71$ to demonstrate that the exponent $\alpha$ is indeed less than unity~\cite{matsuyama2021geometrical}. To this end, we compute several measures of density fluctuations introduced in previous studies, which generalize the structure factor $S(q)$~\cite{torquato2018hyperuniform,berthier2011suppressed,wu2015search, zachary2011hyperuniformity}. 
The first is the correlation function of the local volume fraction
\begin{equation}\label{def1}
S_{1}(q) = \chi_V(q) =\frac{1}{V} \ev{|\phi_1(\mathbf{q})|^2}.
\end{equation}
Here, $\phi_1(\mathbf{q})$ is the Fourier transform of the local volume fraction defined as
\begin{equation}\label{eq:localvolumedef1}
	\phi_1(\mathbf{r}) = \sum_{i} \Delta_i(\mathbf{r} - \mathbf{r}_i),
\end{equation}
where $\Delta_i(\mathbf{r} - \mathbf{r}_i)$ is the indicator function, which equals 1 if $\mathbf{r}$ lies within the $i$-th particle and 0 otherwise.  
The second is the correlation function
\begin{equation}\label{def2}
	S_{2}(q) = \chi(q) = \frac{1}{V} \ev{|\phi_2(\mathbf{q})|^2},
\end{equation}
defined in terms of an alternative local volume fraction $\phi_2(\mathbf{q})$, whose real-space definition is given by
\begin{equation}\label{eq:localvolumedef2}
	\phi_2(\mathbf{r}) = \sum_{i} v_i \delta(\mathbf{r} - \mathbf{r}_i),
\end{equation}
where $v_i = \pi \sigma_i^2 / 4$ is the area of the $i$-th particle. The third measure is the wavenumber-dependent isothermal compressibility, defined for bidisperse systems by
\begin{equation} \label{def3}
S_3(q) = \chi_T(q) 
= \frac{S_{SS}({q}) S_{LL}({q}) - S_{LS}^2({q})}{\frac{N_S^2}{N^2} S_{LL}({q}) + \frac{N_L^2}{N^2} S_{SS}({q}) - 2 \frac{N_S N_L}{N^2} S_{LS}({q})},
\end{equation}
where 
$S_{\nu\mu}(q) = N^{-1} \ev{\delta\rho_\nu(\mathbf{q}) \delta\rho^{\ast}_\mu(\mathbf{q})}$  ($\nu, \mu = S, L$) is the static structure matrix for a binary mixture, 
and $\rho_\nu(\mathbf{q})$ is the density of component
$\nu$~\cite{berthier2011suppressed, matsuyama2021geometrical}.

In Fig.~\ref{fig:threedefinition}, we show all three measures, $S_i(q)$ ($i = 1, 2, 3$),
for $\sr = 0.71$. It is evident that the hyperuniformity exponent
$\alpha$ is not 1 but is instead well fitted by 
$\alpha = 0.6 \text{--} 0.7$ in the intermediate window
$q^{\ast} \sigma_{\text{ave}} \lesssim q \sigma_{\text{ave}} \lesssim 2$, consistent with results reported in Ref.~\cite{matsuyama2021geometrical}. 
We argue below that the exponent is in fact $\alpha = 2/3$. 
Figure~\ref{fig:threedefinition} also demonstrates that, in the hyperuniform regime
(${q^{\ast}  \sigma_{\text{ave}}\lesssim q \sigma_{\text{ave}} \lesssim 2}$), the density 
fluctuations are largely independent of the specific definition of
$S_i(q)$. 

Given this consistency, we use $S_2(q)$ (Eq.~(\ref{def2})) as the reference
structure factor for comparison with our new observable throughout the remainder
of the paper, referring to it simply as the structure factor $S(q)$.

In the following, we also compute the fluctuations of the local volume fraction,
\begin{equation} \label{eq:localvolfluctuation}
\sigma_{\varphi}^2(R) = \ev{\varphi^2(R)} - \ev{\varphi(R)}^2,
\end{equation}
where 
$\varphi(R) = V_R^{-1} \int_{V_R} \phi_1(\mathbf{r}) \, d\mathbf{r}$ 
is
the packing fraction within a circular region of radius $R$ and area $V_R$.  
For hyperuniform systems with $S(q) \propto q^\alpha$ as $q \to 0$, the large-$R$ behavior of $\sigma_\varphi^2(R)$ is~\cite{torquato2018hyperuniform}
\begin{equation}
\sigma_\varphi^2(R) \propto R^{-d - \alpha}, \quad (\alpha < 1).
\label{eq:localvolfluctuation2}
\end{equation}

\section{\label{sec:theory}Generalization of the quasi-particle representation}
\begin{figure}[tb]
	\centering
	\includegraphics[width=0.5\textwidth]{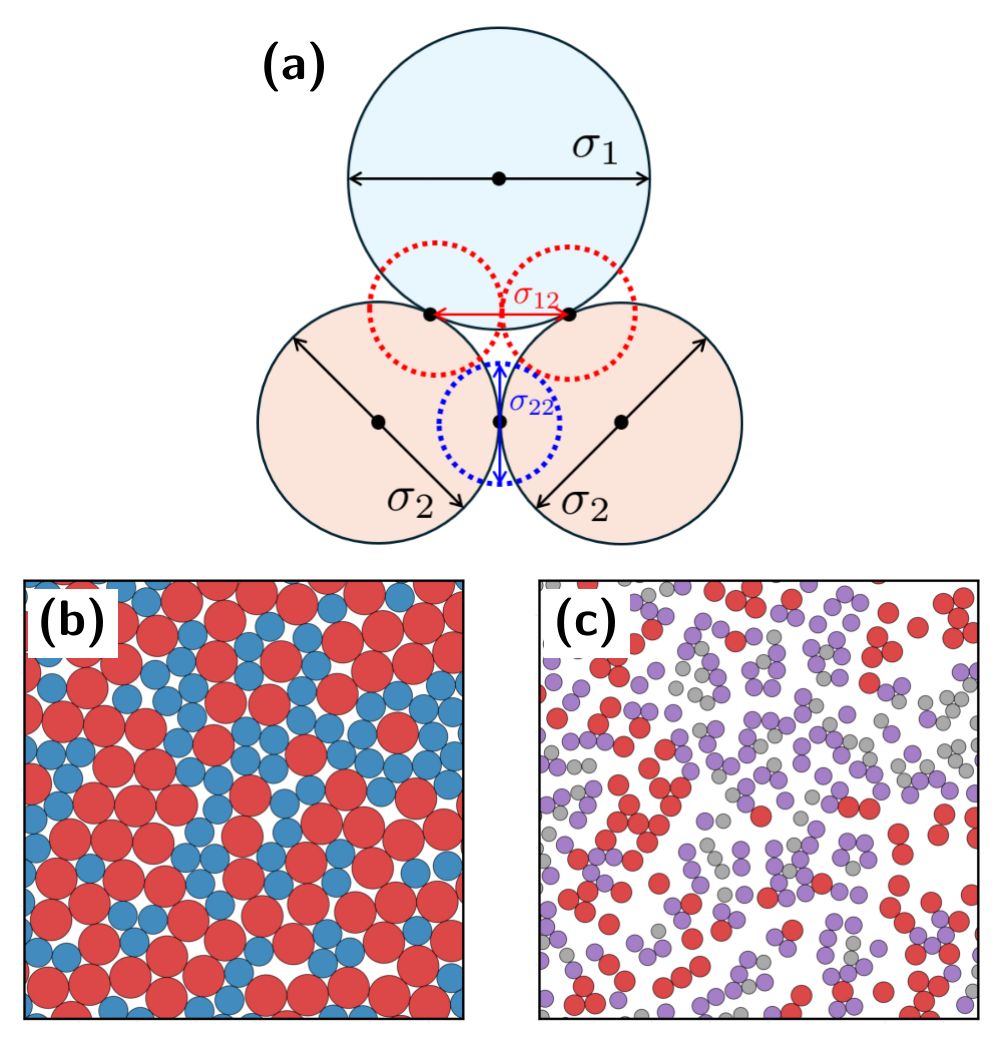}
	\caption{\label{fig:modifiedrissone}
		\justifying
		(a) Schematic of the quasi-particle representation in a bidisperse jammed packing. Three real particles in mutual contact form a locally dense configuration. Quasi-particles (dashed circles) are centered at the contact points, with effective diameters given by Eq.~(\ref{eq:quasiradius}). 
		(b) Portion of a jammed configuration of real particles at size ratio $\sr = 0.71$. 
		(c) Corresponding quasi-particle representation, with quasi-particles located at the contact points of (b). Colors indicate relative particle (effective) diameters.
	}
\end{figure}

Rissone \textit{et al.}~\cite{rissone2021long} recently proposed a method to
compute real-space correlation functions for monodisperse systems 
at long distances with high
accuracy. 
This approach exploits a dual relationship between particle centers
and contact points in isostatic contact networks.
They construct a representation in which contact points are treated as
quasi-particles of effective diameter ${\sigma' = \sigma/2}$, where
$\sigma$ is the particle diameter. This defines the \textit{quasi-particle representation}.
This representation probes correlations within the contact network, which
are structurally distinct from those in the real-particle packing.
Rissone \textit{et al.} analyzed contact--contact correlation functions for observables $\calO$, such as (powers of) the contact number 
and the contact forces, in three-dimensional monodisperse systems.  
They found that the radial correlation function of $\calO$ decays as
$C_\calO(r) \propto r^{-\gamma}$, with an exponent $\gamma \approx 4$. 
This implies $\alpha \approx 1$ for three-dimensional systems, 
as the two exponents are related via the Fourier
transformation by $\alpha = \gamma - d$~\cite{donev2005unexpected}. 
The authors further suggested that such
contact-related observables 
may enhance the precision of long-range correlation measurements,
thereby enabling a more precise determination of the hyperuniformity exponent.

This method is limited to monodisperse systems, where a
rigorous duality exists between contact points and particle centers. 
Although the duality is no longer exact for binary mixtures 
with size disparity, which is necessary to avoid crystallization in $d=2$,
we nevertheless adopt the quasi-particle representation proposed by Rissone et
al., while acknowledging that its theoretical justification is less rigorous in
bidisperse systems. 
In this representation, \textit{quasi-particles} are placed at the contact
points between real particles.  
Figure~\ref{fig:modifiedrissone} provides a schematic illustration of these quasi-particles centered at the points where real particles touch. 
The contact point between the $i$-th and $j$-th particles, denoted by $(ij)$, is
located at 
$\mathbf{r}^c_{ij}=\left(\mathbf{r}_i\sigma_j+\mathbf{r}_j\sigma_i\right)/(\sigma_i+\sigma_j)$.  
To study hyperuniformity in this quasi-particle representation, 
we introduce the local density field defined by 
\begin{equation}
\phi_c(\mathbf{r}) = \sum_{(ij)}\calO_{ij}\delta(\mathbf{r}-\mathbf{r}_{ij}^c),
\label{eq:phic}
\end{equation}
where $\calO_{ij}$ is an observable associated with the $(ij)$-th
quasi-particle, which serves as a weighting variable to improve the visibility
of hyperuniformity~\cite{rissone2021long}.  
In this study, we choose $\calO_{ij}$ to be the local volume defined by
\begin{equation}
v_{ij} = \frac{\pi \left ( \sigma^\mathrm{eff}_{ij} \right )^2}{4},
\label{eq:vij}
\end{equation}
where $\sigma^\mathrm{eff}_{ij}$ is the effective diameter, defined by
\begin{equation}
	\label{eq:quasiradius}
	\frac{1}{\sigma^\mathrm{eff}_{ij}} \equiv \frac{1}{\sigma_i} + \frac{1}{\sigma_j},
\end{equation}
where $\sigma_i$ and $\sigma_j$ denote the diameters of the particles in contact at that quasi-particle.
Our choice of $\calO_{ij}$ is motivated by the geometric interpretation of the jammed packings.  
Equation~(\ref{eq:quasiradius}) assigns a larger effective area to contacts involving larger particles and a smaller area to contacts involving smaller particles.
This choice leads to a more uniform spatial distribution of quasi-particles.
It also appears to mitigate the large compositional fluctuations observed at small $\sr$.
In addition, it yields a high volume fraction while limiting overlap between neighboring quasi-particles,
as illustrated in Fig.~\ref{fig:modifiedrissone} (c) and Fig.~\ref{fig:configplot}.
As $\sr \to 1^{-}$, the results become increasingly insensitive to
the precise form of $\sigma^\mathrm{eff}_{ij}$. 
Using Eqs.~(\ref{eq:phic}) and~(\ref{eq:vij}), we define the structure factor of quasi-particles as
\begin{equation}
S_c(q)=\frac{1}{V}\ev{\left| \sum_{(ij)} v_{ij} e^{-i\mathbf{q}\cdot \mathbf{r}^c_{ij}} \right|^2}.
\label{eq:sqc}
\end{equation}

Our quasi-particle representation is constructed from jammed configurations of real particles.
Although Rissone \textit{et al.}~\cite{rissone2021long} have suggested 
that removing rattlers, defined as particles with fewer than $d+1 = 3$
contacts, is important, we find that the results are insensitive to whether
rattlers are included, at least for $d=2$, as rattlers are either associated with no quasi-particle
($Z=0$), are largely absent ($Z=1$), or occur only rarely as sparse chain-like
structures ($Z=2$) in packings generated by our FIRE energy minimization protocol.
For the range of $\sr$ studied, the generated bidisperse packings are just above the jamming transition density $\varphi_J$ and are nearly isostatic, with an average
contact number per non-rattler particle $\ev{z_i} \approx z_{\text{iso}} = 2d = 4$, except in the monodisperse case, where the system
crystallizes. This yields approximately $2N_{\text{non-rattlers}}$, which is close to $2N$ contact points.
Consequently, the quasi-particle system defined by these contact
points is substantially larger than the original particle system. 

\section{\label{sec:ResultsandDiscussions} Results}

\begin{figure}[tb]
\centering
\includegraphics[width=0.5\textwidth]{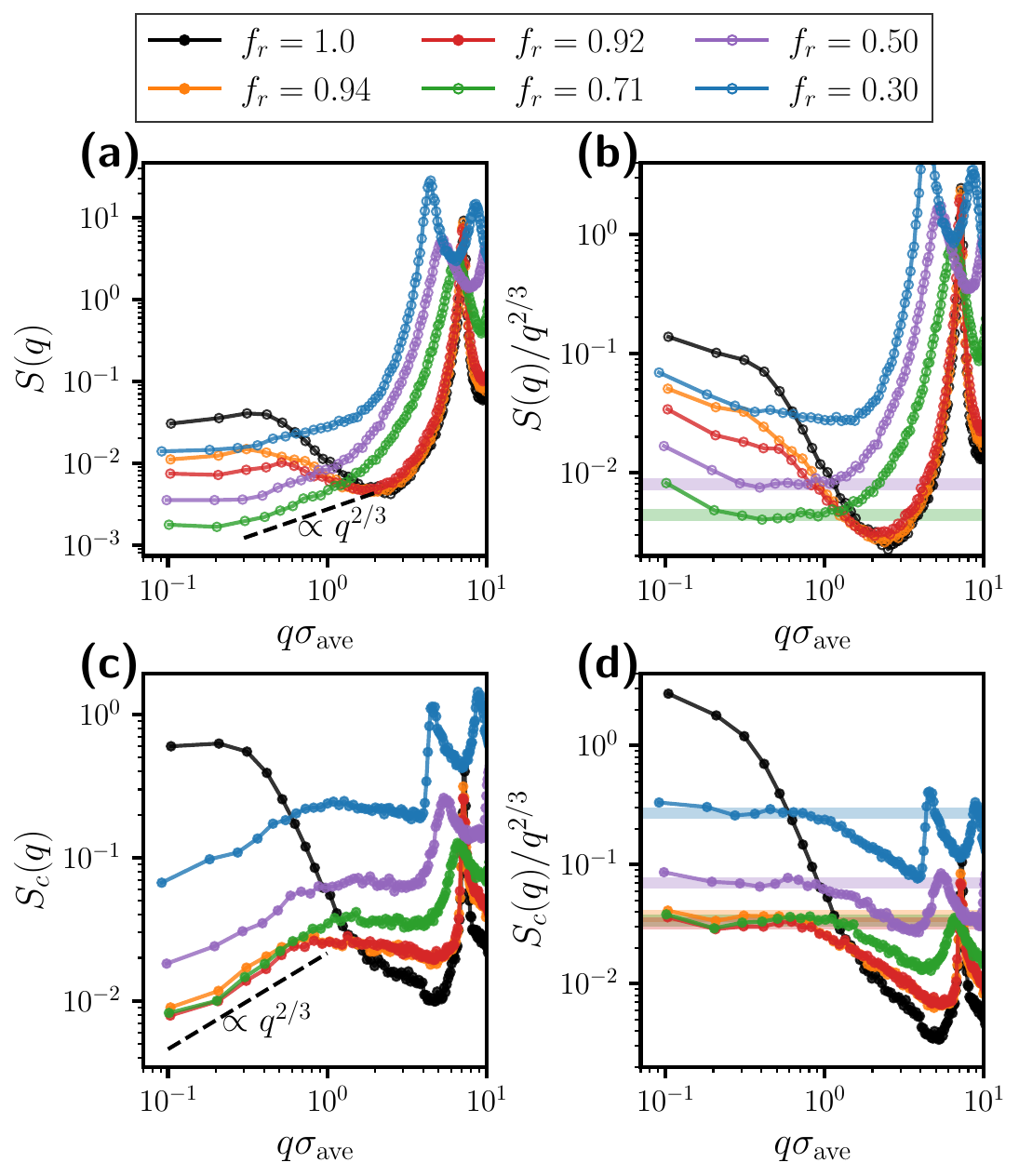}
\caption{\label{fig:combine_chiq}
\justifying  (a) Structure factor $S(q) = S_2(q)$ of the real particles for various size ratios $\sr$. (b) Same data as in (a), rescaled by $q^{2/3}$. 
(c) Structure factor $S_c(q)$ of quasi-particles in the quasi-particle representation for the same $\sr$. (d) Rescaled $S_c(q)/q^{2/3}$, highlighting the low-$q$ plateau. Horizontal lines serve as visual guides.
}
\end{figure}

The jamming transition density $\varphi_J$, contact number, structural
order, and mechanical response vary significantly with the composition
of the mixture, such as the number fraction $\nr$ and size ratio
$\sr$~\cite{koeze2016mapping,saitoh2025jamming,ricouvier2017optimizing}. 
Here, we investigate the dependence of hyperuniformity on the size ratio
$\sr$ over a broad range by
fixing the number fraction at $\nr = 0.5$.

Introducing bidispersity into crystalline monodisperse disk packings
distorts the contact network: contacts are broken, and particles are
displaced from their crystalline positions. These changes, however, are
subtle and often difficult to detect in the real-particle configurations
when $\sr \to 1^{-}$. This suggests that the quasi-particle
representation, which is more sensitive to variations in the contact
network and jamming-related features, provides a more effective
framework for probing long-range density correlations. Indeed, Tong
\textit{et al.}~\cite{tong2015crystals} showed that weak polydispersity
can significantly alter the contact number distribution---and
consequently the distribution of quasi-particles located at the contact points---even though the real particle arrangement remains nearly
crystalline. 
Following the procedure outlined in Sec.~\ref{sec:theory}, we
investigate the structure factor of quasi-particles 
defined by Eq.~(\ref{eq:sqc}), in addition to that of real particles. 

Figure~\ref{fig:combine_chiq} (a) shows the structure factor $S(q)$ of real
particles, defined by Eq.~(\ref{def2}) and measured using the conventional
method, for various size ratios $\sr$. 
For moderate size ratios $\sr=0.5$ and 
$0.71$, 
we observe hyperuniformity with an exponent $\alpha = 0.6\text{--} 0.7$,
for $q >q^{\ast}$, consistent with the $\sr=0.71$ case shown in
Fig.~\ref{fig:threedefinition} using different definitions of $S(q)$.
However, hyperuniform scaling disappears for other values of $\sr$.
In particular, as $\sr$ increases, $S(q)$ exhibits an
enhancement at small wavenumbers $q$. 
The hyperuniform behavior is completely absent for $\sr = 0.92$ (red), $0.94$ (orange), and the monodisperse system $\sr = 1.0$ (black).

\begin{figure}[t]
	\centering
	\includegraphics[width=0.44\textwidth]{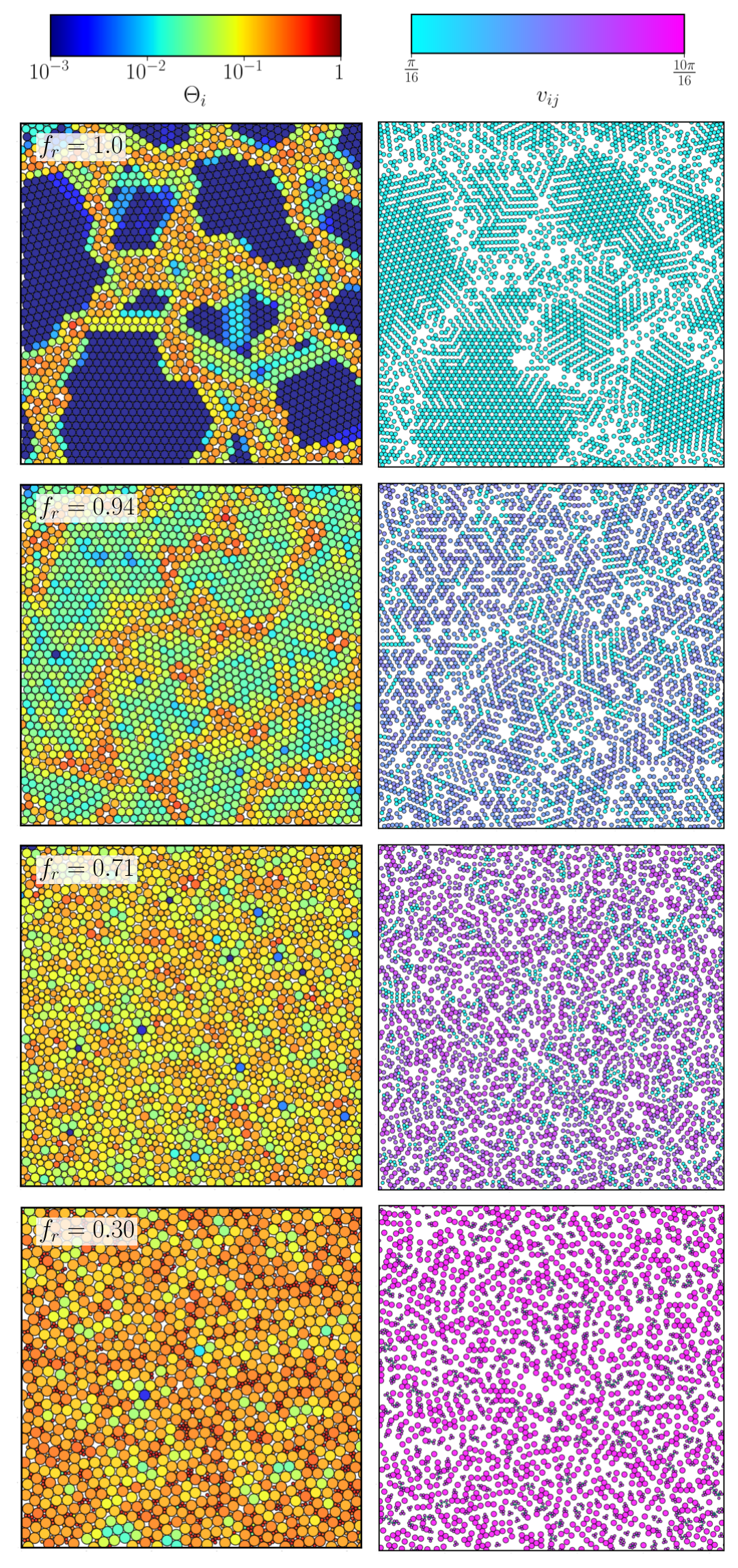}
	\caption{\justifying
Snapshots of selected regions of jammed configurations. The size ratio $\sr$ decreases from top ($\sr = 1.0$) to bottom ($\sr = 0.3$). 
Left: real particles, colored by the Tong--Tanaka structural order parameter~\cite{tong2019structural}. 
		Right: quasi-particles, colored by the volume $v_{ij} = {\pi \left(\sigma ^\mathrm{eff}_{ij} \right) ^2/4}$ defined in Eq.~(\ref{eq:vij}).
		Real particles and quasi-particles are shown with sizes set by their diameters and effective diameters, respectively.
	}
	\label{fig:configplot}
\end{figure}

Figure~\ref{fig:configplot} presents selected regions of jammed packings
for several values of $\sr$.  
The left panels show configurations of real particles, with colors representing the Tong--Tanaka local packing order parameter for each particle~\cite{tong2019structural}. 
This order parameter for the $i$-th particle is defined as
\begin{equation}
	\Theta_i = \frac{1}{N_i} \sum_{\langle jk \rangle} \left| \theta_{jk}^{(1)} - \theta_{jk}^{(2)} \right|,
\end{equation}
where $N_i$ is the number of neighboring particle pairs $\langle jk\rangle$ adjacent to the $i$-th particle, $\theta_{jk}^{(1)}$ is the
angle between ${\mathbf{r}_{ij}=\mathbf{r}_i-\mathbf{r}_j}$ and ${\mathbf{r}_{ik}=\mathbf{r}_i-\mathbf{r}_k}$ for each
such pair, and $\theta_{jk}^{(2)}$ is the corresponding angle when the
three particles are perfectly in contact.
This parameter quantifies
local packing efficiency: particles shown in blue are more densely packed, whereas those in red are more loosely packed. The right panels show the corresponding quasi-particle configurations derived from the contact networks of the left panels. Here, the colors represent the quasi-particle volumes, computed from Eq.~(\ref{eq:vij}), relative to the smallest quasi-particle. From the configurations of real particles, we observe that packings with moderate size ratios such as $\sr = 0.71$ exhibit no evident structural order and are not efficiently packed. Consistent with this, these systems are expected to exhibit the hyperuniformity typical of disordered jammed packings, as
confirmed in Fig.~\ref{fig:combine_chiq} (a). 
In contrast, for size ratios approaching the monodisperse limit ($\sr \to 1^{-}$), such as $\sr = 0.94$, and in the monodisperse case ($\sr = 1.0$), particles form crystalline domains. The emergence of polycrystalline domain boundaries and fluctuations in domain sizes lead to an enhancement in the small-wavenumber limit, as reflected in the behavior of $S(q)$.

Note that jamming criticality, such as 
${\delta z \propto \delta\varphi^{1/2}}$, persists even in the nearly
monodisperse regime (${\sr \to 1^{-}}$)~\cite{tong2015crystals,ikeda2020jamming}. 
The enhancement of
$S(q)$ at large $\sr$ (i.e., as $\sr \to 1^{-}$), which masks the underlying hyperuniformity, is attributed to the formation of grain boundaries associated with
crystalline ordering, as noted above. The typical size of
polycrystalline domains can be estimated from the structure factors in
Fig.~\ref{fig:combine_chiq} (a). 
The domain size is given by $L_d=2\pi/q_d$, where $q_d$ denotes the crossover wavenumber below which $S(q)$ exhibits an enhancement. These estimates are consistent with the structural domain sizes observed in Fig.~\ref{fig:configplot}. 
For instance, we find $L_d \approx
8\sigma_{\text{ave}}$ for $\sr = 0.94$. 

\begin{figure}[tb]
	\centering
	\includegraphics[width=0.5\textwidth]{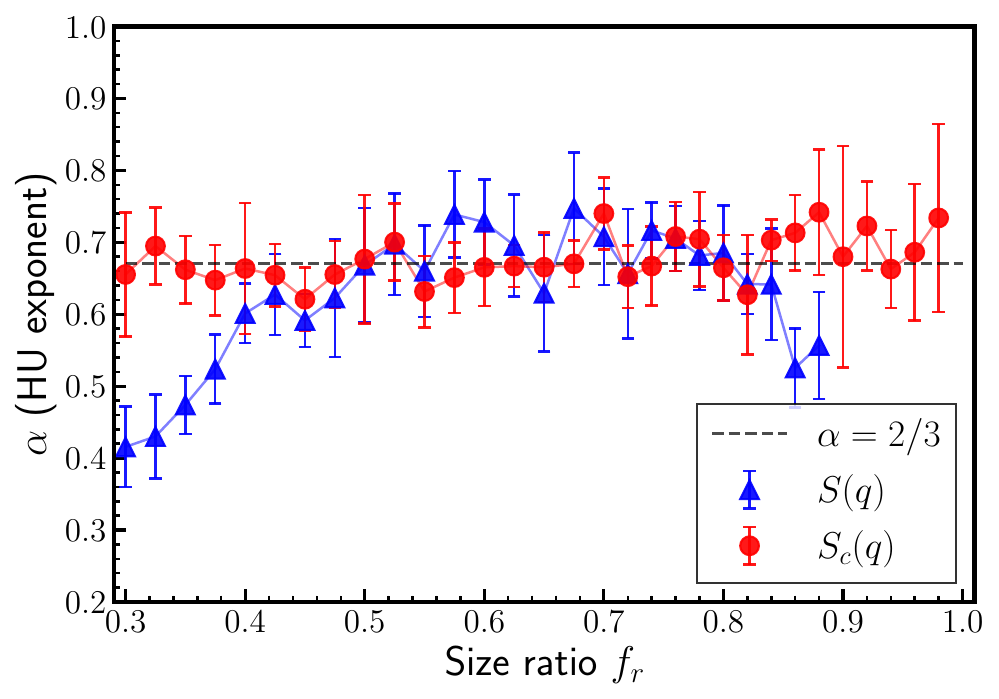}
	\caption{\label{fig:alphaplot} \justifying 
	The hyperuniformity exponent $\alpha$ measured from the structure factor of the real particles 
(Eq.~(\ref{def2}), blue) and of the quasi-particles (Eq.~(\ref{eq:sqc}), red). 
The dashed line indicates the predicted value $\alpha = 2/3$.
}
\end{figure}

Let us now turn our attention to the quasi-particle representation. 
Figure~\ref{fig:combine_chiq} (c) shows the structure factor $S_c(q)$
defined by Eq.~(\ref{eq:sqc}) for quasi-particles in this
representation. Remarkably, $S_c(q)$ decays toward zero as $q
\to 0$ for all $\sr$ in the range $0.3 \leq \sr \leq 0.94$,
following a power law $S_c(q) \propto q^{\alpha}$ with $\alpha \approx 2/3$, indicating robust hyperuniformity. To highlight the
improved accuracy of $S_c(q)$ relative to $S(q)$, we plot both $S(q)$
and $S_c(q)$ rescaled by $q^{2/3}$ in Figs.~\ref{fig:combine_chiq} (b)
and \ref{fig:combine_chiq} (d), respectively. 
Both quantities exhibit a plateau at small $q$ for moderate size ratios $0.5 \leq \sr \leq 0.71$. 
The deviation from the plateau is observed for $q < q^{\ast}$, suggesting that the crossover wavenumber $q^{\ast}$ is weakly dependent on $\sr$.
For systems near the monodisperse limit ($\sr > 0.92$) and for strongly
size-disparate systems ($\sr = 0.3$), $S(q)/q^{2/3}$ shows significant enhancement in the small-wavenumber limit, deviating from the expected scaling. 
In contrast, $S_c(q)/q^{2/3}$ maintains a plateau across the entire range of
$\sr$ studied, except for the monodisperse system $\sr = 1.0$.  
Furthermore, the $q$ window over which the plateau is observed is consistently broader for $S_c(q)$ than for $S(q)$. 
To further assess this improvement, we verify that our choice of quasi-particle
observable maximizes the visibility of density hyperuniformity, as detailed in
Appendix~\ref{appendix:weightingexponent}. 
From these observations, we conclude that the quasi-particle analysis provides a
more robust characterization compared to the real particle representation,  
with hyperuniformity in the intermediate-$q$ regime above $q^{\ast}$  
persisting robustly up to nearly the monodisperse limit. 
The monodisperse packing ($\sr = 1.0$) remains an exception, where both $S(q)$ and $S_c(q)$ consistently show enhancement in the small-wavenumber limit, indicating a finite domain
size of $L_d \approx 15\sigma_{\text{ave}}$. 

To quantify hyperuniformity across different size ratios $\sr$, we fit
the power-law form in Eq.~(\ref{eq:HUasymtotic_S}) to the structure
factors $S(q)$ of real particles and $S_c(q)$ of quasi-particles.
Because the exponent $\alpha$ is sensitive to the fitting range, we select
fitting ranges of $q$ for which variations in the range have minimal effects on
$\alpha$~\cite{maher2023hyperuniformity}.    
Figure~\ref{fig:alphaplot} shows the values of $\alpha$ obtained from both
$S(q)$ and $S_c(q)$ over the entire range $0.3 \leq \sr \leq 1$.   
The $\alpha$ values derived from $S(q)$ exhibit a non-monotonic dependence on
$\sr$: they start as low as 0.4, peak at $\alpha \approx 0.7$ for moderate size
ratios ($0.5 \leq \sr \leq 0.8$) and then decrease as $\sr$ approaches the
monodisperse limit at 1.   
In contrast, the $\alpha$ values extracted from $S_c(q)$ remain nearly constant, fluctuating around $2/3$ throughout the entire range of $\sr$.
We also examined the system with a different number fraction 
$\nr = N_S / (N_S + N_L) = 0.2$ (data not shown)
and found that the result remains unchanged,  i.e., 
$\alpha \approx 0.6\text{--}0.7$ for $S_c(q)$ for the entire range of $\sr$.

\begin{figure}[tb]
\centering
\includegraphics[width=0.45\textwidth]{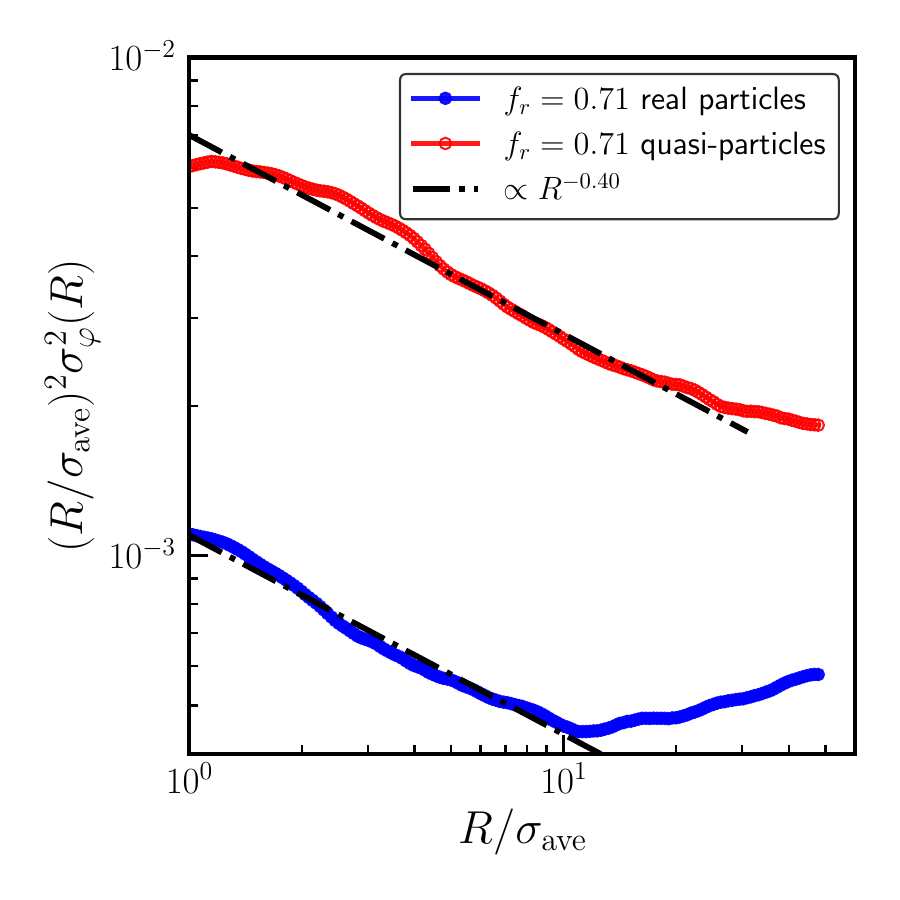}
\caption {\label{fig:volumefluctuation_N1e4_main}  
\justifying 
{
Local volume fluctuation $\sigma_\varphi^2(R)$, using definition 1 of the
 local volume fraction (Eq. \ref{eq:localvolumedef1}),  for the disordered
 jammed system with $\sr=0.71$ and $N=10\,000$.}
} 
\end{figure}

Finally, we investigate real-space fluctuations of the local volume
fraction for both real particles and quasi-particles.
As shown in Ref.~\cite{wu2015search}, although the large-scale density
fluctuations inferred from Fourier space are independent of the specific
definition of the local volume fraction, the real-space fluctuations
$\sigma_\varphi^2(R)$ can differ substantially between definitions. In
particular, the contribution to $\sigma_\varphi^2(R)$ from intermediate wavenumbers is significantly larger when using definition~2
(Eq.~(\ref{eq:localvolumedef2})) than when using definition~1
(Eq.~(\ref{eq:localvolumedef1})). 
Consequently, the large-$R$ behavior of
$\sigma_\varphi^2(R)$ obtained from definition~1 is expected to more faithfully reflect the corresponding Fourier-space behavior. For this reason, we compute the
real-space local volume fluctuations using definition~1. 
In systems with a hyperuniformity exponent $\alpha < 1$, the real-space
signature of hyperuniformity is the anomalous suppression of local volume
fraction fluctuations, characterized by the scaling relation 
$R^d \sigma_\varphi^2(R) \propto R^{-\alpha}$~\cite{torquato2018hyperuniform}. 
To
suppress statistical noise and mitigate finite-size effects, we adopt the method
of Dreyfus \textit{et al.}~\cite{dreyfus2015diagnosing}, in which the original
configuration is replicated in a $3 \times 3$ tiling to construct a larger
fictitious system. This stitched configuration remains jammed due to periodic
boundary conditions.  
Figure~\ref{fig:volumefluctuation_N1e4_main} shows the local volume-fraction
fluctuations for the size ratio $\sr = 1/1.4 \approx 0.71$ and system size $N =
10^4$. For the real particles (blue dots), the rescaled fluctuations
$R^{2}\sigma^2_{\varphi}(R)$ decay approximately as $R^{-0.4}$ from small $R$ up
to a crossover length $R^{*} \approx 10\,\sigma_{\mathrm{ave}}$.  
We note that $2\pi/(2R^{*})\approx 0.3 \sigma_{\mathrm{ave}}^{-1}$ is roughly consistent
with the crossover wavenumber $q^{\ast}$ below which hyperuniform scaling in the 
Fourier-space representation breaks down. 
However, the exponent extracted from the real-space data is ${\alpha_{\mathrm{real}}\approx 0.4}$,
which is appreciably smaller than $2/3$ obtained from $S(q)$ and $S_c(q)$.
For the quasi-particles (red dots), we also
observe the $R^{-0.4}$ scaling, which appears to persist to length scales
larger than $2R^{\ast}$.  
The physical origin of the distinct exponent $0.4$ is unclear. 
This exponent is identical to the results of the real-space measurements reported in
previous work~\cite{wu2015search}, which also show similar inconsistencies
between the real- and Fourier-space measurements. 
Taken together with the significant dependence of $\sigma_\varphi^2(R)$ 
on the definition of the local volume fraction, these observations suggest that
the Fourier-space representations, $S(q)$ and $S_c(q)$, for both real particles
and quasi-particles provide a more reliable and robust characterization of hyperuniformity.

\section{Conclusions and Discussion}\label{sec:conclusionanddiscussion} 

In this work, we investigated hyperuniformity in two-dimensional jammed
packings, focusing on its dependence on bidispersity. Hyperuniformity is
characterized by a vanishing structure factor, $S(q) \propto q^{\alpha}$
with $\alpha > 0$ as $q \to 0$. We examined this behavior both in the
spatial distribution of real particles and in that of quasi-particles
centered at contact points. Across a wide range of size ratios, we consistently
observe signature of hyperuniformity with an exponent $\alpha = 0.6\text{--}0.7$, 
down to a crossover wavenumber 
$q^{\ast}\approx 0.2\sigma_{\mathrm{ave}}^{-1}$, 
in agreement with our previous results for the
classic $1\!:\!1.4$ equimolar binary
mixture~\cite{matsuyama2021geometrical}. 
Hyperuniformity is more
clearly observed in the quasi-particle representation, particularly near
the monodisperse limit where local crystalline order in real particle
configurations masks long-range correlations. In contrast, the quasi-particle framework reveals robust hyperuniform behavior with $\alpha \approx 2/3$ across all bidisperse systems, except for the monodisperse limit corresponding to the ideal triangular crystal. These findings underscore the importance of
the contact-based approach in capturing long-range structural features
and demonstrate its utility for probing hyperuniformity in jammed
systems beyond the reach of conventional particle-based analyses. 

In hindsight, the exponent $\alpha = 2/3$ is not surprising, given the mean-field nature of the jamming transition~\cite{matsuyama2021geometrical}.
For an ideally hyperuniform system, the asymptotic Fourier-space behavior $S(q) \propto q^\alpha$ implies the large-$R$ scaling $\sigma_\varphi^2(R) \propto R^{-d-\alpha}$ for $\alpha < 1$ in real space.
On the other hand, given that the upper critical dimension of the jamming
transition is $d_c = 2$~\cite{Wyart2005,goodrich2012finite}, mean-field theory
suggests that the fluctuation $\sigma_\varphi^2(R)$ scales with the system size $N$ as 
$\sigma_\varphi^2(R) \propto N^{-\Omega}$~\cite{Binder1985prb,Wittmann2014pre}.  
Combining $N \propto R^d$ with the above hyperuniform scaling for $\sigma_\varphi^2(R)$, 
one expects $\alpha$ to scale proportionally with dimension $d$ as $\alpha = d(\Omega-1)$. 
If $\alpha=1$ for $d=3$, then one obtains $\Omega=4/3$. This implies the hyperuniformity exponent $\alpha=2/3$ for $d=2$.

We emphasize that the hyperuniformity discussed in this study is density
hyperuniformity and that the windows over which the algebraic law holds are
limited to ${q \gtrsim q^{\ast}}$. 
However, the existence of the power law and its associated exponent appears robust and universal.
This observation of hyperuniformity in an intermediate wavenumber window is distinct from the hyperuniformity with $\alpha \approx 0.45$ reported by Maher \textit{et al.}~\cite{maher2024hyperuniformity} for MRJ-like binary disk packings, where the exponent was extracted
from large-$R$ volume fraction fluctuations. This contrast shows that hyperuniformity in jammed systems depends on both the preparation protocol and the probing length scale, an issue that remains unresolved.
Recently, two distinct forms of hyperuniformity have attracted much attention near the jamming
transition point, $\varphi_{J}$.
One is density hyperuniformity that intrinsically originates from the absorbing transition~\cite{wilken2021random, wilken2023dynamical,Wang2025d,wang2025hyperuniform}.  
Another is the hyperuniformity of the excess contact number, which is
accompanied by diverging length scales near the jamming
transition~\cite{hexner2018two,hexner2019can}.  
How these forms of hyperuniformity are related to the finite-length-scale
behaviors studied here remains an open question, warranting further investigation. 

\begin{acknowledgments}
This work was supported by KAKENHI (Grant Numbers 
JP20H00128, 
JP22H04472, 
JP23H04503, 
JP24H00192),   
and the JST FOREST Program (Grant Number JPMJFR212T). 
\end{acknowledgments}

\appendix

\section{\label{appendix:weightingexponent} Effect of weighting variable on quasi-particle hyperuniformity}
\renewcommand{\theequation}{\Alph{section}\arabic{equation}}  
\renewcommand{\thefigure}{\Alph{section}\arabic{figure}}  
\setcounter{equation}{0}
\setcounter{figure}{0}

\begin{figure}[t]
	\centering
	\includegraphics[width=0.5\textwidth]{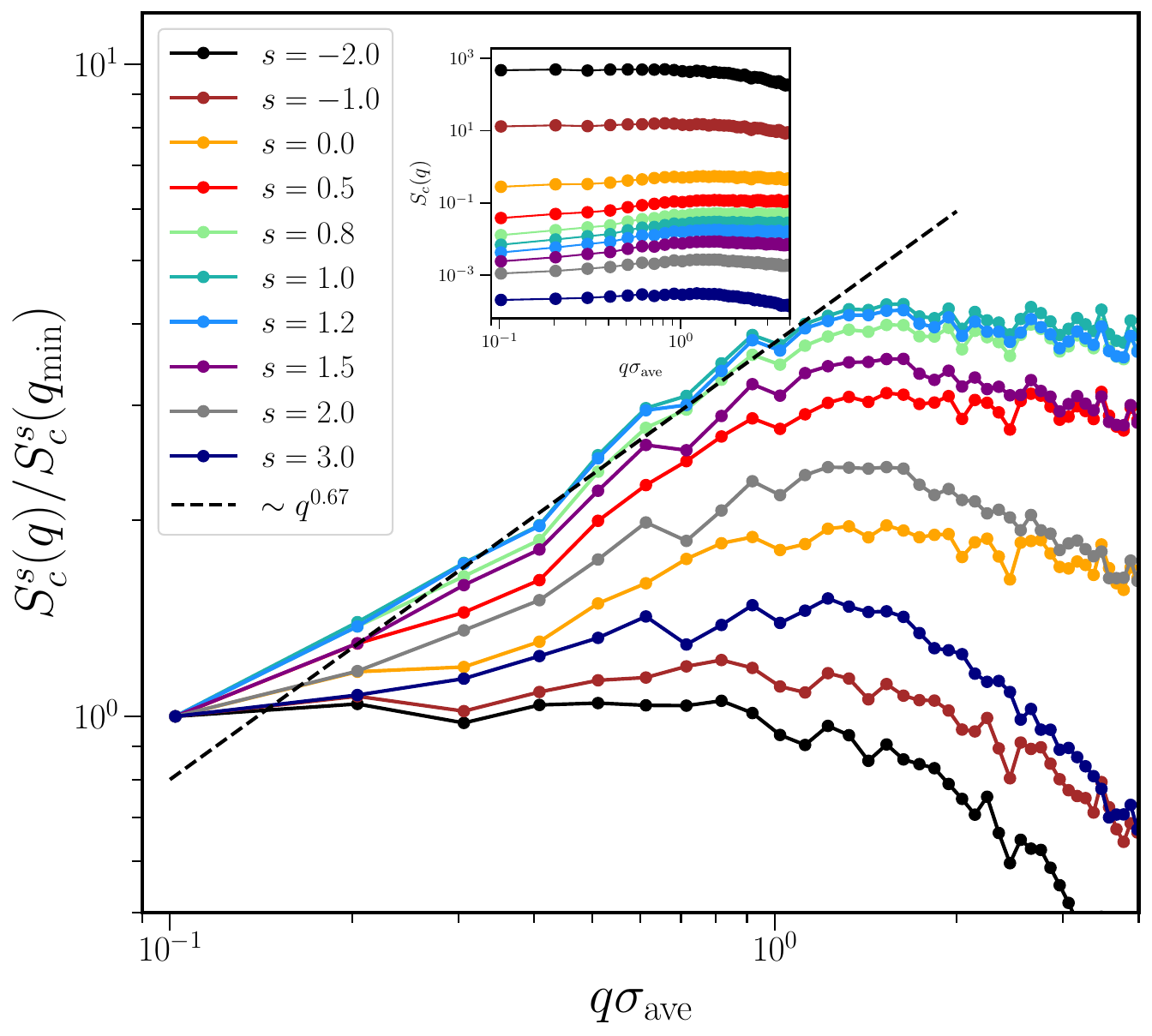}
	\caption{\label{fig:general_Sc}
		\justifying
		Structure factor of quasi-particles calculated for different exponents $s$, as defined in Eq.~(\ref{eq:ssc}). The vertical axis is rescaled by the value of each structure factor at the smallest accessible wavenumber, $S^s_c(q_{\text{min}})$. The size ratio is $f_r=0.71$. The inset shows the same data without rescaling.
	}
\end{figure}

In this section, we demonstrate the role of the weighting variables $\calO_{ij}$ on the degree of hyperuniformity observed in the quasi-particle representation. Following Rissone \textit{et al.}~\cite{rissone2021long}, the local density field associated with an observable $\mathcal{O}_{ij}$ (Eq.~\eqref{eq:phic}) can be generalized as
\begin{equation}
	\label{eq:phisc}
	\phi^s_c(\mathbf{r}) = \sum_{(ij)}\calO^s_{ij}\delta(\mathbf{r}-\mathbf{r}_{ij}^c),
\end{equation}
where the exponent $s$ adjusts the weighting of the observable, thereby tuning its contribution to long-range correlations. For instance, when $\mathcal{O}_{ij}$ corresponds to the contact force, Rissone \textit{et al.} found that $s=-1/2$ emphasizes weak contact forces, and that the resulting correlation function exhibits hyperuniformity with exponent $\alpha=1$, as noted above.

Following this framework, we show that the choice $\mathcal{O}_{ij}=v_{ij}$ (Eq.~\eqref{eq:vij}) with $s=1$ is appropriate for revealing the long-range hyperuniform behavior in our system. To this end, we examine the generalized structure factor, defined in Fourier space as
\begin{equation}
	S^s_c(q) = \frac{1}{V}\ev{\left|\sum_{(ij)} v_{ij}^{s} e^{i{\bf q}\cdot{\bf r}^c_{ij}}\right|^2},
	\label{eq:ssc}
\end{equation}
for several values of the parameter $s$. Here, $s=0$ corresponds to the bare density correlation function of the quasi-particles, $s=1$ corresponds to the correlation function used in this paper (Eq.~(\ref{eq:sqc})), i.e., the structure factor constructed from the field weighted by $v_{ij}$, and $s=0.5$ corresponds to weighting by the effective diameter, since $v_{ij} \propto \left( \sigma^\mathrm{eff}_{ij}\right)^2$ implies $\mathcal{O}_{ij}^{0.5} \propto \sigma^\mathrm{eff}_{ij}$.

Figure~\ref{fig:general_Sc} shows $S^s_c(q)$ for $s$ ranging from $-2$ to $3$, 
at size ratio $f_r=0.71$. For negative $s$, no suppression of fluctuations at 
small-$q$ is observed. As $s$ increases toward positive values, signatures of hyperuniformity begin to emerge for ${q \gtrsim q^{\ast}}$. The hyperuniform behavior is most pronounced at $s=1$, beyond which it weakens with increasing $s$. At $s=1$, we recover the hyperuniform behavior shown in Fig.~\ref{fig:combine_chiq}, with hyperuniformity exponent 
$\alpha= 0.6\text{--}0.7$. At present, we do not have a clear explanation for 
why $s=1$ is the optimal value.

\section{\label{ap:systemsize} System size dependence}

\begin{figure}[t]
	\centering
	\includegraphics[width=0.5\textwidth]{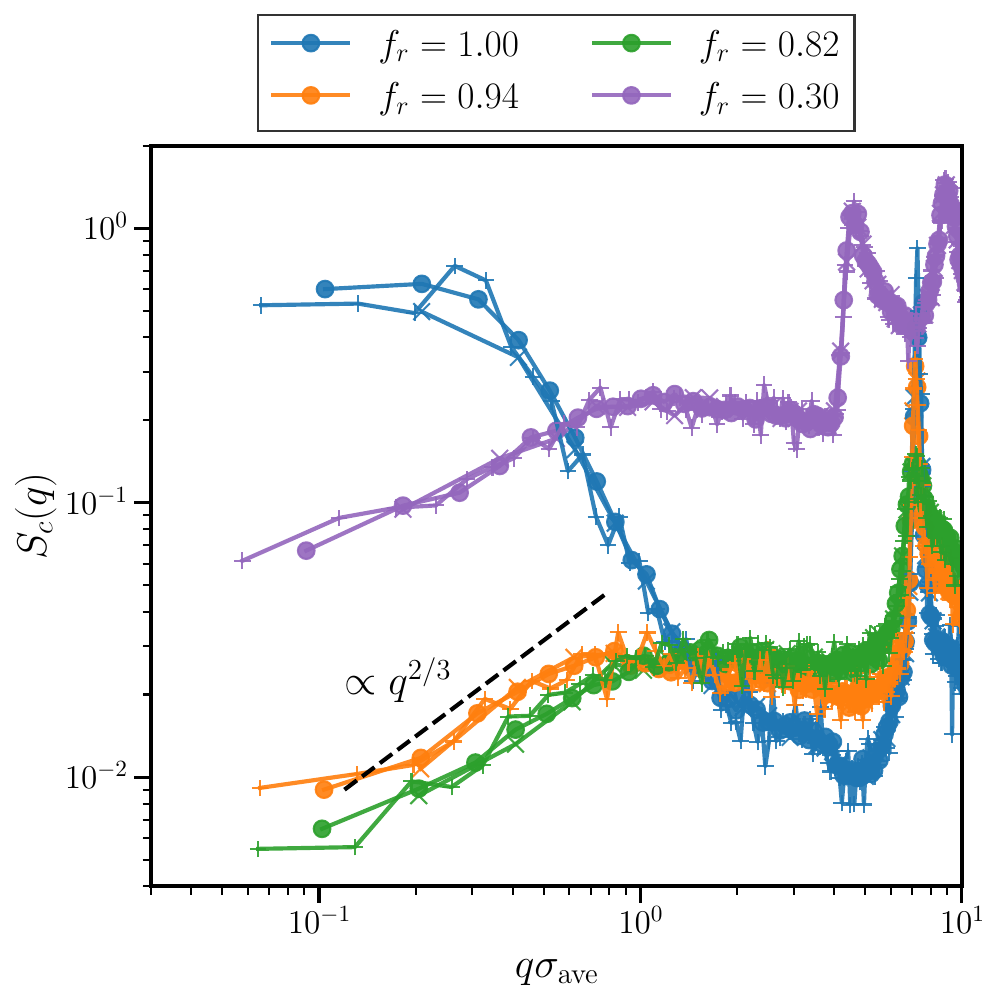}
	\caption{\label{fig:systemsize} \justifying The structure factor $S_c(q)$ of the quasi-particles for systems of size $N=1\,000$ ($\times$), $N=4\,000$ ($\circ$), and $N=10\,000$ ($+$), for an equimolar bidisperse system with various size ratios $\sr$.}
\end{figure}

Figure~\ref{fig:systemsize} shows the structure factor of quasi-particles for systems of different sizes, $N = 1\,000$, $4\,000$, and $10\,000$, for several size ratios $\sr$ in an equimolar bidisperse system. Due to periodic boundary conditions, larger systems provide access to smaller wavenumbers, with $q_{\min} \propto 1/\sqrt{N}$. We observe that the structure factors collapse onto a single curve, indicating that the hyperuniformity measured in the quasi-particle representation is robust with respect to system size.

\begin{thebibliography}{49}%
\makeatletter
\providecommand \@ifxundefined [1]{%
 \@ifx{#1\undefined}
}%
\providecommand \@ifnum [1]{%
 \ifnum #1\expandafter \@firstoftwo
 \else \expandafter \@secondoftwo
 \fi
}%
\providecommand \@ifx [1]{%
 \ifx #1\expandafter \@firstoftwo
 \else \expandafter \@secondoftwo
 \fi
}%
\providecommand \natexlab [1]{#1}%
\providecommand \enquote  [1]{``#1''}%
\providecommand \bibnamefont  [1]{#1}%
\providecommand \bibfnamefont [1]{#1}%
\providecommand \citenamefont [1]{#1}%
\providecommand \href@noop [0]{\@secondoftwo}%
\providecommand \href [0]{\begingroup \@sanitize@url \@href}%
\providecommand \@href[1]{\@@startlink{#1}\@@href}%
\providecommand \@@href[1]{\endgroup#1\@@endlink}%
\providecommand \@sanitize@url [0]{\catcode `\\12\catcode `\$12\catcode
  `\&12\catcode `\#12\catcode `\^12\catcode `\_12\catcode `\%12\relax}%
\providecommand \@@startlink[1]{}%
\providecommand \@@endlink[0]{}%
\providecommand \url  [0]{\begingroup\@sanitize@url \@url }%
\providecommand \@url [1]{\endgroup\@href {#1}{\urlprefix }}%
\providecommand \urlprefix  [0]{URL }%
\providecommand \Eprint [0]{\href }%
\providecommand \doibase [0]{https://doi.org/}%
\providecommand \selectlanguage [0]{\@gobble}%
\providecommand \bibinfo  [0]{\@secondoftwo}%
\providecommand \bibfield  [0]{\@secondoftwo}%
\providecommand \translation [1]{[#1]}%
\providecommand \BibitemOpen [0]{}%
\providecommand \bibitemStop [0]{}%
\providecommand \bibitemNoStop [0]{.\EOS\space}%
\providecommand \EOS [0]{\spacefactor3000\relax}%
\providecommand \BibitemShut  [1]{\csname bibitem#1\endcsname}%
\let\auto@bib@innerbib\@empty
\bibitem [{\citenamefont {Torquato}\ and\ \citenamefont
  {Stillinger}(2003)}]{torquato2003local}%
  \BibitemOpen
  \bibfield  {author} {\bibinfo {author} {\bibfnamefont {S.}~\bibnamefont
  {Torquato}}\ and\ \bibinfo {author} {\bibfnamefont {F.~H.}\ \bibnamefont
  {Stillinger}},\ }\bibfield  {title} {\bibinfo {title} {Local density
  fluctuations, hyperuniformity, and order metrics},\ }\href
  {https://doi.org/10.1103/PhysRevE.68.041113} {\bibfield  {journal} {\bibinfo
  {journal} {Phys. Rev. E}\ }\textbf {\bibinfo {volume} {68}},\ \bibinfo
  {pages} {041113} (\bibinfo {year} {2003})}\BibitemShut {NoStop}%
\bibitem [{\citenamefont {Torquato}(2018)}]{torquato2018hyperuniform}%
  \BibitemOpen
  \bibfield  {author} {\bibinfo {author} {\bibfnamefont {S.}~\bibnamefont
  {Torquato}},\ }\bibfield  {title} {\bibinfo {title} {Hyperuniform states of
  matter},\ }\href
  {https://doi.org/https://doi.org/10.1016/j.physrep.2018.03.001} {\bibfield
  {journal} {\bibinfo  {journal} {Physics Reports}\ }\textbf {\bibinfo {volume}
  {745}},\ \bibinfo {pages} {1} (\bibinfo {year} {2018})}\BibitemShut {NoStop}%
\bibitem [{\citenamefont {Gabrielli}\ \emph {et~al.}(2002)\citenamefont
  {Gabrielli}, \citenamefont {Joyce},\ and\ \citenamefont
  {Sylos~Labini}}]{gabrielli2002glass}%
  \BibitemOpen
  \bibfield  {author} {\bibinfo {author} {\bibfnamefont {A.}~\bibnamefont
  {Gabrielli}}, \bibinfo {author} {\bibfnamefont {M.}~\bibnamefont {Joyce}},\
  and\ \bibinfo {author} {\bibfnamefont {F.}~\bibnamefont {Sylos~Labini}},\
  }\bibfield  {title} {\bibinfo {title} {Glass-like universe: Real-space
  correlation properties of standard cosmological models},\ }\href
  {https://doi.org/10.1103/PhysRevD.65.083523} {\bibfield  {journal} {\bibinfo
  {journal} {Phys. Rev. D}\ }\textbf {\bibinfo {volume} {65}},\ \bibinfo
  {pages} {083523} (\bibinfo {year} {2002})}\BibitemShut {NoStop}%
\bibitem [{\citenamefont {Jiao}\ \emph {et~al.}(2014)\citenamefont {Jiao},
  \citenamefont {Lau}, \citenamefont {Hatzikirou}, \citenamefont
  {Meyer-Hermann}, \citenamefont {Corbo},\ and\ \citenamefont
  {Torquato}}]{jiao2014avian}%
  \BibitemOpen
  \bibfield  {author} {\bibinfo {author} {\bibfnamefont {Y.}~\bibnamefont
  {Jiao}}, \bibinfo {author} {\bibfnamefont {T.}~\bibnamefont {Lau}}, \bibinfo
  {author} {\bibfnamefont {H.}~\bibnamefont {Hatzikirou}}, \bibinfo {author}
  {\bibfnamefont {M.}~\bibnamefont {Meyer-Hermann}}, \bibinfo {author}
  {\bibfnamefont {J.~C.}\ \bibnamefont {Corbo}},\ and\ \bibinfo {author}
  {\bibfnamefont {S.}~\bibnamefont {Torquato}},\ }\bibfield  {title} {\bibinfo
  {title} {Avian photoreceptor patterns represent a disordered hyperuniform
  solution to a multiscale packing problem},\ }\href
  {https://doi.org/10.1103/PhysRevE.89.022721} {\bibfield  {journal} {\bibinfo
  {journal} {Phys. Rev. E}\ }\textbf {\bibinfo {volume} {89}},\ \bibinfo
  {pages} {022721} (\bibinfo {year} {2014})}\BibitemShut {NoStop}%
\bibitem [{\citenamefont {Mayer}\ \emph {et~al.}(2015)\citenamefont {Mayer},
  \citenamefont {Balasubramanian}, \citenamefont {Mora},\ and\ \citenamefont
  {Walczak}}]{mayer2015well}%
  \BibitemOpen
  \bibfield  {author} {\bibinfo {author} {\bibfnamefont {A.}~\bibnamefont
  {Mayer}}, \bibinfo {author} {\bibfnamefont {V.}~\bibnamefont
  {Balasubramanian}}, \bibinfo {author} {\bibfnamefont {T.}~\bibnamefont
  {Mora}},\ and\ \bibinfo {author} {\bibfnamefont {A.~M.}\ \bibnamefont
  {Walczak}},\ }\bibfield  {title} {\bibinfo {title} {How a well-adapted immune
  system is organized},\ }\href {https://doi.org/10.1073/pnas.1421827112}
  {\bibfield  {journal} {\bibinfo  {journal} {Proceedings of the National
  Academy of Sciences}\ }\textbf {\bibinfo {volume} {112}},\ \bibinfo {pages}
  {5950} (\bibinfo {year} {2015})},\ \Eprint
  {https://arxiv.org/abs/https://www.pnas.org/doi/pdf/10.1073/pnas.1421827112}
  {https://www.pnas.org/doi/pdf/10.1073/pnas.1421827112} \BibitemShut {NoStop}%
\bibitem [{\citenamefont {Zhang}\ \emph {et~al.}(2016)\citenamefont {Zhang},
  \citenamefont {Stillinger},\ and\ \citenamefont
  {Torquato}}]{zhang2016perfect}%
  \BibitemOpen
  \bibfield  {author} {\bibinfo {author} {\bibfnamefont {G.}~\bibnamefont
  {Zhang}}, \bibinfo {author} {\bibfnamefont {F.~H.}\ \bibnamefont
  {Stillinger}},\ and\ \bibinfo {author} {\bibfnamefont {S.}~\bibnamefont
  {Torquato}},\ }\bibfield  {title} {\bibinfo {title} {The perfect glass
  paradigm: Disordered hyperuniform glasses down to absolute zero},\ }\href
  {https://doi.org/10.1038/srep36963} {\bibfield  {journal} {\bibinfo
  {journal} {Scientific Reports}\ }\textbf {\bibinfo {volume} {6}},\ \bibinfo
  {pages} {36963} (\bibinfo {year} {2016})}\BibitemShut {NoStop}%
\bibitem [{\citenamefont {Donev}\ \emph
  {et~al.}(2005{\natexlab{a}})\citenamefont {Donev}, \citenamefont {Torquato},\
  and\ \citenamefont {Stillinger}}]{donev2005pair}%
  \BibitemOpen
  \bibfield  {author} {\bibinfo {author} {\bibfnamefont {A.}~\bibnamefont
  {Donev}}, \bibinfo {author} {\bibfnamefont {S.}~\bibnamefont {Torquato}},\
  and\ \bibinfo {author} {\bibfnamefont {F.~H.}\ \bibnamefont {Stillinger}},\
  }\bibfield  {title} {\bibinfo {title} {Pair correlation function
  characteristics of nearly jammed disordered and ordered hard-sphere
  packings},\ }\href {https://doi.org/10.1103/PhysRevE.71.011105} {\bibfield
  {journal} {\bibinfo  {journal} {Phys. Rev. E}\ }\textbf {\bibinfo {volume}
  {71}},\ \bibinfo {pages} {011105} (\bibinfo {year}
  {2005}{\natexlab{a}})}\BibitemShut {NoStop}%
\bibitem [{\citenamefont {Zachary}\ \emph
  {et~al.}(2011{\natexlab{a}})\citenamefont {Zachary}, \citenamefont {Jiao},\
  and\ \citenamefont {Torquato}}]{zachary2011hyperuniform}%
  \BibitemOpen
  \bibfield  {author} {\bibinfo {author} {\bibfnamefont {C.~E.}\ \bibnamefont
  {Zachary}}, \bibinfo {author} {\bibfnamefont {Y.}~\bibnamefont {Jiao}},\ and\
  \bibinfo {author} {\bibfnamefont {S.}~\bibnamefont {Torquato}},\ }\bibfield
  {title} {\bibinfo {title} {Hyperuniform long-range correlations are a
  signature of disordered jammed hard-particle packings},\ }\href
  {https://doi.org/10.1103/PhysRevLett.106.178001} {\bibfield  {journal}
  {\bibinfo  {journal} {Phys. Rev. Lett.}\ }\textbf {\bibinfo {volume} {106}},\
  \bibinfo {pages} {178001} (\bibinfo {year} {2011}{\natexlab{a}})}\BibitemShut
  {NoStop}%
\bibitem [{\citenamefont {Zachary}\ \emph
  {et~al.}(2011{\natexlab{b}})\citenamefont {Zachary}, \citenamefont {Jiao},\
  and\ \citenamefont {Torquato}}]{zachary2011hyperuniformity}%
  \BibitemOpen
  \bibfield  {author} {\bibinfo {author} {\bibfnamefont {C.~E.}\ \bibnamefont
  {Zachary}}, \bibinfo {author} {\bibfnamefont {Y.}~\bibnamefont {Jiao}},\ and\
  \bibinfo {author} {\bibfnamefont {S.}~\bibnamefont {Torquato}},\ }\bibfield
  {title} {\bibinfo {title} {Hyperuniformity, quasi-long-range correlations,
  and void-space constraints in maximally random jammed particle packings. {I.
  Polydisperse spheres}},\ }\href {https://doi.org/10.1103/PhysRevE.83.051308}
  {\bibfield  {journal} {\bibinfo  {journal} {Phys. Rev. E}\ }\textbf {\bibinfo
  {volume} {83}},\ \bibinfo {pages} {051308} (\bibinfo {year}
  {2011}{\natexlab{b}})}\BibitemShut {NoStop}%
\bibitem [{\citenamefont {Atkinson}\ \emph {et~al.}(2016)\citenamefont
  {Atkinson}, \citenamefont {Zhang}, \citenamefont {Hopkins},\ and\
  \citenamefont {Torquato}}]{atkinson2016critical}%
  \BibitemOpen
  \bibfield  {author} {\bibinfo {author} {\bibfnamefont {S.}~\bibnamefont
  {Atkinson}}, \bibinfo {author} {\bibfnamefont {G.}~\bibnamefont {Zhang}},
  \bibinfo {author} {\bibfnamefont {A.~B.}\ \bibnamefont {Hopkins}},\ and\
  \bibinfo {author} {\bibfnamefont {S.}~\bibnamefont {Torquato}},\ }\bibfield
  {title} {\bibinfo {title} {Critical slowing down and hyperuniformity on
  approach to jamming},\ }\href {https://doi.org/10.1103/PhysRevE.94.012902}
  {\bibfield  {journal} {\bibinfo  {journal} {Phys. Rev. E}\ }\textbf {\bibinfo
  {volume} {94}},\ \bibinfo {pages} {012902} (\bibinfo {year}
  {2016})}\BibitemShut {NoStop}%
\bibitem [{\citenamefont {Liu}\ and\ \citenamefont
  {Nagel}(1998)}]{liu1998jamming}%
  \BibitemOpen
  \bibfield  {author} {\bibinfo {author} {\bibfnamefont {A.~J.}\ \bibnamefont
  {Liu}}\ and\ \bibinfo {author} {\bibfnamefont {S.~R.}\ \bibnamefont
  {Nagel}},\ }\bibfield  {title} {\bibinfo {title} {Jamming is not just cool
  any more},\ }\href {https://doi.org/10.1038/23819} {\bibfield  {journal}
  {\bibinfo  {journal} {Nature}\ }\textbf {\bibinfo {volume} {396}},\ \bibinfo
  {pages} {21} (\bibinfo {year} {1998})}\BibitemShut {NoStop}%
\bibitem [{\citenamefont {O'Hern}\ \emph {et~al.}(2003)\citenamefont {O'Hern},
  \citenamefont {Silbert}, \citenamefont {Liu},\ and\ \citenamefont
  {Nagel}}]{o2003jamming}%
  \BibitemOpen
  \bibfield  {author} {\bibinfo {author} {\bibfnamefont {C.~S.}\ \bibnamefont
  {O'Hern}}, \bibinfo {author} {\bibfnamefont {L.~E.}\ \bibnamefont {Silbert}},
  \bibinfo {author} {\bibfnamefont {A.~J.}\ \bibnamefont {Liu}},\ and\ \bibinfo
  {author} {\bibfnamefont {S.~R.}\ \bibnamefont {Nagel}},\ }\bibfield  {title}
  {\bibinfo {title} {Jamming at zero temperature and zero applied stress: The
  epitome of disorder},\ }\href {https://doi.org/10.1103/PhysRevE.68.011306}
  {\bibfield  {journal} {\bibinfo  {journal} {Phys. Rev. E}\ }\textbf {\bibinfo
  {volume} {68}},\ \bibinfo {pages} {011306} (\bibinfo {year}
  {2003})}\BibitemShut {NoStop}%
\bibitem [{\citenamefont {van Hecke}(2009)}]{van2009jamming}%
  \BibitemOpen
  \bibfield  {author} {\bibinfo {author} {\bibfnamefont {M.}~\bibnamefont {van
  Hecke}},\ }\bibfield  {title} {\bibinfo {title} {Jamming of soft particles:
  geometry, mechanics, scaling and isostaticity},\ }\href
  {https://doi.org/10.1088/0953-8984/22/3/033101} {\bibfield  {journal}
  {\bibinfo  {journal} {Journal of Physics: Condensed Matter}\ }\textbf
  {\bibinfo {volume} {22}},\ \bibinfo {pages} {033101} (\bibinfo {year}
  {2009})}\BibitemShut {NoStop}%
\bibitem [{\citenamefont {{Wyart, M.}}(2005)}]{Wyart2005}%
  \BibitemOpen
  \bibfield  {author} {\bibinfo {author} {\bibnamefont {{Wyart, M.}}},\
  }\bibfield  {title} {\bibinfo {title} {On the rigidity of amorphous solids},\
  }\href {https://doi.org/10.1051/anphys:2006003} {\bibfield  {journal}
  {\bibinfo  {journal} {Ann. Phys. Fr.}\ }\textbf {\bibinfo {volume} {30}},\
  \bibinfo {pages} {1} (\bibinfo {year} {2005})}\BibitemShut {NoStop}%
\bibitem [{\citenamefont {Charbonneau}\ \emph {et~al.}(2014)\citenamefont
  {Charbonneau}, \citenamefont {Kurchan}, \citenamefont {Parisi}, \citenamefont
  {Urbani},\ and\ \citenamefont {Zamponi}}]{charbonneau2014fractal}%
  \BibitemOpen
  \bibfield  {author} {\bibinfo {author} {\bibfnamefont {P.}~\bibnamefont
  {Charbonneau}}, \bibinfo {author} {\bibfnamefont {J.}~\bibnamefont
  {Kurchan}}, \bibinfo {author} {\bibfnamefont {G.}~\bibnamefont {Parisi}},
  \bibinfo {author} {\bibfnamefont {P.}~\bibnamefont {Urbani}},\ and\ \bibinfo
  {author} {\bibfnamefont {F.}~\bibnamefont {Zamponi}},\ }\bibfield  {title}
  {\bibinfo {title} {Fractal free energy landscapes in structural glasses},\
  }\href {https://doi.org/10.1038/ncomms4725} {\bibfield  {journal} {\bibinfo
  {journal} {Nature Communications}\ }\textbf {\bibinfo {volume} {5}},\
  \bibinfo {pages} {3725} (\bibinfo {year} {2014})}\BibitemShut {NoStop}%
\bibitem [{\citenamefont {Torquato}\ \emph {et~al.}(2000)\citenamefont
  {Torquato}, \citenamefont {Truskett},\ and\ \citenamefont
  {Debenedetti}}]{torquato2000random}%
  \BibitemOpen
  \bibfield  {author} {\bibinfo {author} {\bibfnamefont {S.}~\bibnamefont
  {Torquato}}, \bibinfo {author} {\bibfnamefont {T.~M.}\ \bibnamefont
  {Truskett}},\ and\ \bibinfo {author} {\bibfnamefont {P.~G.}\ \bibnamefont
  {Debenedetti}},\ }\bibfield  {title} {\bibinfo {title} {Is random close
  packing of spheres well defined?},\ }\href
  {https://doi.org/10.1103/PhysRevLett.84.2064} {\bibfield  {journal} {\bibinfo
   {journal} {Phys. Rev. Lett.}\ }\textbf {\bibinfo {volume} {84}},\ \bibinfo
  {pages} {2064} (\bibinfo {year} {2000})}\BibitemShut {NoStop}%
\bibitem [{\citenamefont {Donev}\ \emph
  {et~al.}(2005{\natexlab{b}})\citenamefont {Donev}, \citenamefont
  {Stillinger},\ and\ \citenamefont {Torquato}}]{donev2005unexpected}%
  \BibitemOpen
  \bibfield  {author} {\bibinfo {author} {\bibfnamefont {A.}~\bibnamefont
  {Donev}}, \bibinfo {author} {\bibfnamefont {F.~H.}\ \bibnamefont
  {Stillinger}},\ and\ \bibinfo {author} {\bibfnamefont {S.}~\bibnamefont
  {Torquato}},\ }\bibfield  {title} {\bibinfo {title} {Unexpected density
  fluctuations in jammed disordered sphere packings},\ }\href
  {https://doi.org/10.1103/PhysRevLett.95.090604} {\bibfield  {journal}
  {\bibinfo  {journal} {Phys. Rev. Lett.}\ }\textbf {\bibinfo {volume} {95}},\
  \bibinfo {pages} {090604} (\bibinfo {year} {2005}{\natexlab{b}})}\BibitemShut
  {NoStop}%
\bibitem [{\citenamefont {Berthier}\ \emph {et~al.}(2011)\citenamefont
  {Berthier}, \citenamefont {Chaudhuri}, \citenamefont {Coulais}, \citenamefont
  {Dauchot},\ and\ \citenamefont {Sollich}}]{berthier2011suppressed}%
  \BibitemOpen
  \bibfield  {author} {\bibinfo {author} {\bibfnamefont {L.}~\bibnamefont
  {Berthier}}, \bibinfo {author} {\bibfnamefont {P.}~\bibnamefont {Chaudhuri}},
  \bibinfo {author} {\bibfnamefont {C.}~\bibnamefont {Coulais}}, \bibinfo
  {author} {\bibfnamefont {O.}~\bibnamefont {Dauchot}},\ and\ \bibinfo {author}
  {\bibfnamefont {P.}~\bibnamefont {Sollich}},\ }\bibfield  {title} {\bibinfo
  {title} {Suppressed compressibility at large scale in jammed packings of
  size-disperse spheres},\ }\href
  {https://doi.org/10.1103/PhysRevLett.106.120601} {\bibfield  {journal}
  {\bibinfo  {journal} {Phys. Rev. Lett.}\ }\textbf {\bibinfo {volume} {106}},\
  \bibinfo {pages} {120601} (\bibinfo {year} {2011})}\BibitemShut {NoStop}%
\bibitem [{\citenamefont {Wu}\ \emph {et~al.}(2015)\citenamefont {Wu},
  \citenamefont {Olsson},\ and\ \citenamefont {Teitel}}]{wu2015search}%
  \BibitemOpen
  \bibfield  {author} {\bibinfo {author} {\bibfnamefont {Y.}~\bibnamefont
  {Wu}}, \bibinfo {author} {\bibfnamefont {P.}~\bibnamefont {Olsson}},\ and\
  \bibinfo {author} {\bibfnamefont {S.}~\bibnamefont {Teitel}},\ }\bibfield
  {title} {\bibinfo {title} {Search for hyperuniformity in mechanically stable
  packings of frictionless disks above jamming},\ }\href
  {https://doi.org/10.1103/PhysRevE.92.052206} {\bibfield  {journal} {\bibinfo
  {journal} {Phys. Rev. E}\ }\textbf {\bibinfo {volume} {92}},\ \bibinfo
  {pages} {052206} (\bibinfo {year} {2015})}\BibitemShut {NoStop}%
\bibitem [{\citenamefont {Dreyfus}\ \emph {et~al.}(2015)\citenamefont
  {Dreyfus}, \citenamefont {Xu}, \citenamefont {Still}, \citenamefont {Hough},
  \citenamefont {Yodh},\ and\ \citenamefont
  {Torquato}}]{dreyfus2015diagnosing}%
  \BibitemOpen
  \bibfield  {author} {\bibinfo {author} {\bibfnamefont {R.}~\bibnamefont
  {Dreyfus}}, \bibinfo {author} {\bibfnamefont {Y.}~\bibnamefont {Xu}},
  \bibinfo {author} {\bibfnamefont {T.}~\bibnamefont {Still}}, \bibinfo
  {author} {\bibfnamefont {L.~A.}\ \bibnamefont {Hough}}, \bibinfo {author}
  {\bibfnamefont {A.~G.}\ \bibnamefont {Yodh}},\ and\ \bibinfo {author}
  {\bibfnamefont {S.}~\bibnamefont {Torquato}},\ }\bibfield  {title} {\bibinfo
  {title} {Diagnosing hyperuniformity in two-dimensional, disordered, jammed
  packings of soft spheres},\ }\href
  {https://doi.org/10.1103/PhysRevE.91.012302} {\bibfield  {journal} {\bibinfo
  {journal} {Phys. Rev. E}\ }\textbf {\bibinfo {volume} {91}},\ \bibinfo
  {pages} {012302} (\bibinfo {year} {2015})}\BibitemShut {NoStop}%
\bibitem [{\citenamefont {Ozawa}\ \emph {et~al.}(2017)\citenamefont {Ozawa},
  \citenamefont {Berthier},\ and\ \citenamefont
  {Coslovich}}]{ozawa2017exploring}%
  \BibitemOpen
  \bibfield  {author} {\bibinfo {author} {\bibfnamefont {M.}~\bibnamefont
  {Ozawa}}, \bibinfo {author} {\bibfnamefont {L.}~\bibnamefont {Berthier}},\
  and\ \bibinfo {author} {\bibfnamefont {D.}~\bibnamefont {Coslovich}},\
  }\bibfield  {title} {\bibinfo {title} {{Exploring the jamming transition over
  a wide range of critical densities}},\ }\href
  {https://doi.org/10.21468/SciPostPhys.3.4.027} {\bibfield  {journal}
  {\bibinfo  {journal} {SciPost Phys.}\ }\textbf {\bibinfo {volume} {3}},\
  \bibinfo {pages} {027} (\bibinfo {year} {2017})}\BibitemShut {NoStop}%
\bibitem [{\citenamefont {Chieco}\ \emph {et~al.}(2018)\citenamefont {Chieco},
  \citenamefont {Zu}, \citenamefont {Liu}, \citenamefont {Xu},\ and\
  \citenamefont {Durian}}]{chieco2018spectrum}%
  \BibitemOpen
  \bibfield  {author} {\bibinfo {author} {\bibfnamefont {A.~T.}\ \bibnamefont
  {Chieco}}, \bibinfo {author} {\bibfnamefont {M.}~\bibnamefont {Zu}}, \bibinfo
  {author} {\bibfnamefont {A.~J.}\ \bibnamefont {Liu}}, \bibinfo {author}
  {\bibfnamefont {N.}~\bibnamefont {Xu}},\ and\ \bibinfo {author}
  {\bibfnamefont {D.~J.}\ \bibnamefont {Durian}},\ }\bibfield  {title}
  {\bibinfo {title} {Spectrum of structure for jammed and unjammed soft
  disks},\ }\href {https://doi.org/10.1103/PhysRevE.98.042606} {\bibfield
  {journal} {\bibinfo  {journal} {Phys. Rev. E}\ }\textbf {\bibinfo {volume}
  {98}},\ \bibinfo {pages} {042606} (\bibinfo {year} {2018})}\BibitemShut
  {NoStop}%
\bibitem [{\citenamefont {Ikeda}\ \emph {et~al.}(2017)\citenamefont {Ikeda},
  \citenamefont {Berthier},\ and\ \citenamefont {Parisi}}]{ikeda2017large}%
  \BibitemOpen
  \bibfield  {author} {\bibinfo {author} {\bibfnamefont {A.}~\bibnamefont
  {Ikeda}}, \bibinfo {author} {\bibfnamefont {L.}~\bibnamefont {Berthier}},\
  and\ \bibinfo {author} {\bibfnamefont {G.}~\bibnamefont {Parisi}},\
  }\bibfield  {title} {\bibinfo {title} {Large-scale structure of randomly
  jammed spheres},\ }\href {https://doi.org/10.1103/PhysRevE.95.052125}
  {\bibfield  {journal} {\bibinfo  {journal} {Phys. Rev. E}\ }\textbf {\bibinfo
  {volume} {95}},\ \bibinfo {pages} {052125} (\bibinfo {year}
  {2017})}\BibitemShut {NoStop}%
\bibitem [{\citenamefont {Ikeda}\ and\ \citenamefont
  {Berthier}(2015)}]{ikeda2015thermal}%
  \BibitemOpen
  \bibfield  {author} {\bibinfo {author} {\bibfnamefont {A.}~\bibnamefont
  {Ikeda}}\ and\ \bibinfo {author} {\bibfnamefont {L.}~\bibnamefont
  {Berthier}},\ }\bibfield  {title} {\bibinfo {title} {Thermal fluctuations,
  mechanical response, and hyperuniformity in jammed solids},\ }\href
  {https://doi.org/10.1103/PhysRevE.92.012309} {\bibfield  {journal} {\bibinfo
  {journal} {Phys. Rev. E}\ }\textbf {\bibinfo {volume} {92}},\ \bibinfo
  {pages} {012309} (\bibinfo {year} {2015})}\BibitemShut {NoStop}%
\bibitem [{\citenamefont {Xu}\ and\ \citenamefont
  {Ching}(2010)}]{xu2010effects}%
  \BibitemOpen
  \bibfield  {author} {\bibinfo {author} {\bibfnamefont {N.}~\bibnamefont
  {Xu}}\ and\ \bibinfo {author} {\bibfnamefont {E.~S.~C.}\ \bibnamefont
  {Ching}},\ }\bibfield  {title} {\bibinfo {title} {Effects of particle-size
  ratio on jamming of binary mixtures at zero temperature},\ }\href
  {https://doi.org/10.1039/B926696H} {\bibfield  {journal} {\bibinfo  {journal}
  {Soft Matter}\ }\textbf {\bibinfo {volume} {6}},\ \bibinfo {pages} {2944}
  (\bibinfo {year} {2010})}\BibitemShut {NoStop}%
\bibitem [{\citenamefont {Henkel}\ \emph {et~al.}(2008)\citenamefont {Henkel},
  \citenamefont {Hinrichsen},\ and\ \citenamefont
  {L{\"{u}}beck}}]{Henkel2008book}%
  \BibitemOpen
  \bibfield  {author} {\bibinfo {author} {\bibfnamefont {M.}~\bibnamefont
  {Henkel}}, \bibinfo {author} {\bibfnamefont {H.}~\bibnamefont {Hinrichsen}},\
  and\ \bibinfo {author} {\bibfnamefont {S.}~\bibnamefont {L{\"{u}}beck}},\
  }\href {https://doi.org/10.1007/978-1-4020-8765-3} {\emph {\bibinfo {title}
  {{"Non-Equilibrium Phase Transitions Volume 1: Absorbing Phase
  Transitions"}}}}\ (\bibinfo  {publisher} {Springer},\ \bibinfo {address}
  {Dordrecht},\ \bibinfo {year} {2008})\BibitemShut {NoStop}%
\bibitem [{\citenamefont {Wilken}\ \emph {et~al.}(2021)\citenamefont {Wilken},
  \citenamefont {Guerra}, \citenamefont {Levine},\ and\ \citenamefont
  {Chaikin}}]{wilken2021random}%
  \BibitemOpen
  \bibfield  {author} {\bibinfo {author} {\bibfnamefont {S.}~\bibnamefont
  {Wilken}}, \bibinfo {author} {\bibfnamefont {R.~E.}\ \bibnamefont {Guerra}},
  \bibinfo {author} {\bibfnamefont {D.}~\bibnamefont {Levine}},\ and\ \bibinfo
  {author} {\bibfnamefont {P.~M.}\ \bibnamefont {Chaikin}},\ }\bibfield
  {title} {\bibinfo {title} {Random close packing as a dynamical phase
  transition},\ }\href {https://doi.org/10.1103/PhysRevLett.127.038002}
  {\bibfield  {journal} {\bibinfo  {journal} {Phys. Rev. Lett.}\ }\textbf
  {\bibinfo {volume} {127}},\ \bibinfo {pages} {038002} (\bibinfo {year}
  {2021})}\BibitemShut {NoStop}%
\bibitem [{\citenamefont {Wilken}\ \emph {et~al.}(2023)\citenamefont {Wilken},
  \citenamefont {Guo}, \citenamefont {Levine},\ and\ \citenamefont
  {Chaikin}}]{wilken2023dynamical}%
  \BibitemOpen
  \bibfield  {author} {\bibinfo {author} {\bibfnamefont {S.}~\bibnamefont
  {Wilken}}, \bibinfo {author} {\bibfnamefont {A.~Z.}\ \bibnamefont {Guo}},
  \bibinfo {author} {\bibfnamefont {D.}~\bibnamefont {Levine}},\ and\ \bibinfo
  {author} {\bibfnamefont {P.~M.}\ \bibnamefont {Chaikin}},\ }\bibfield
  {title} {\bibinfo {title} {Dynamical approach to the jamming problem},\
  }\href {https://doi.org/10.1103/PhysRevLett.131.238202} {\bibfield  {journal}
  {\bibinfo  {journal} {Phys. Rev. Lett.}\ }\textbf {\bibinfo {volume} {131}},\
  \bibinfo {pages} {238202} (\bibinfo {year} {2023})}\BibitemShut {NoStop}%
\bibitem [{\citenamefont {Matsuyama}\ \emph {et~al.}(2021)\citenamefont
  {Matsuyama}, \citenamefont {Toyoda}, \citenamefont {Kurahashi}, \citenamefont
  {Ikeda}, \citenamefont {Kawasaki},\ and\ \citenamefont
  {Miyazaki}}]{matsuyama2021geometrical}%
  \BibitemOpen
  \bibfield  {author} {\bibinfo {author} {\bibfnamefont {H.}~\bibnamefont
  {Matsuyama}}, \bibinfo {author} {\bibfnamefont {M.}~\bibnamefont {Toyoda}},
  \bibinfo {author} {\bibfnamefont {T.}~\bibnamefont {Kurahashi}}, \bibinfo
  {author} {\bibfnamefont {A.}~\bibnamefont {Ikeda}}, \bibinfo {author}
  {\bibfnamefont {T.}~\bibnamefont {Kawasaki}},\ and\ \bibinfo {author}
  {\bibfnamefont {K.}~\bibnamefont {Miyazaki}},\ }\bibfield  {title} {\bibinfo
  {title} {Geometrical properties of mechanically annealed systems near the
  jamming transition},\ }\href
  {https://doi.org/10.1140/epje/s10189-021-00142-6} {\bibfield  {journal}
  {\bibinfo  {journal} {The European Physical Journal E}\ }\textbf {\bibinfo
  {volume} {44}},\ \bibinfo {pages} {133} (\bibinfo {year} {2021})}\BibitemShut
  {NoStop}%
\bibitem [{\citenamefont {Koeze}\ \emph {et~al.}(2016)\citenamefont {Koeze},
  \citenamefont {V^^c3^^a5gberg}, \citenamefont {Tjoa},\ and\ \citenamefont
  {Tighe}}]{koeze2016mapping}%
  \BibitemOpen
  \bibfield  {author} {\bibinfo {author} {\bibfnamefont {D.~J.}\ \bibnamefont
  {Koeze}}, \bibinfo {author} {\bibfnamefont {D.}~\bibnamefont
  {V^^c3^^a5gberg}}, \bibinfo {author} {\bibfnamefont {B.~B.~T.}\ \bibnamefont
  {Tjoa}},\ and\ \bibinfo {author} {\bibfnamefont {B.~P.}\ \bibnamefont
  {Tighe}},\ }\bibfield  {title} {\bibinfo {title} {Mapping the jamming
  transition of bidisperse mixtures},\ }\href
  {https://doi.org/10.1209/0295-5075/113/54001} {\bibfield  {journal} {\bibinfo
   {journal} {Europhysics Letters}\ }\textbf {\bibinfo {volume} {113}},\
  \bibinfo {pages} {54001} (\bibinfo {year} {2016})}\BibitemShut {NoStop}%
\bibitem [{\citenamefont {Saitoh}\ and\ \citenamefont
  {Tighe}(2025)}]{saitoh2025jamming}%
  \BibitemOpen
  \bibfield  {author} {\bibinfo {author} {\bibfnamefont {K.}~\bibnamefont
  {Saitoh}}\ and\ \bibinfo {author} {\bibfnamefont {B.~P.}\ \bibnamefont
  {Tighe}},\ }\bibfield  {title} {\bibinfo {title} {Jamming transition and
  normal modes of polydispersed soft particle packing},\ }\href
  {https://doi.org/10.1039/D4SM01305K} {\bibfield  {journal} {\bibinfo
  {journal} {Soft Matter}\ }\textbf {\bibinfo {volume} {21}},\ \bibinfo {pages}
  {1263} (\bibinfo {year} {2025})}\BibitemShut {NoStop}%
\bibitem [{\citenamefont {Ricouvier}\ \emph {et~al.}(2017)\citenamefont
  {Ricouvier}, \citenamefont {Pierrat}, \citenamefont {Carminati},
  \citenamefont {Tabeling},\ and\ \citenamefont
  {Yazhgur}}]{ricouvier2017optimizing}%
  \BibitemOpen
  \bibfield  {author} {\bibinfo {author} {\bibfnamefont {J.}~\bibnamefont
  {Ricouvier}}, \bibinfo {author} {\bibfnamefont {R.}~\bibnamefont {Pierrat}},
  \bibinfo {author} {\bibfnamefont {R.}~\bibnamefont {Carminati}}, \bibinfo
  {author} {\bibfnamefont {P.}~\bibnamefont {Tabeling}},\ and\ \bibinfo
  {author} {\bibfnamefont {P.}~\bibnamefont {Yazhgur}},\ }\bibfield  {title}
  {\bibinfo {title} {Optimizing hyperuniformity in self-assembled bidisperse
  emulsions},\ }\href {https://doi.org/10.1103/PhysRevLett.119.208001}
  {\bibfield  {journal} {\bibinfo  {journal} {Phys. Rev. Lett.}\ }\textbf
  {\bibinfo {volume} {119}},\ \bibinfo {pages} {208001} (\bibinfo {year}
  {2017})}\BibitemShut {NoStop}%
\bibitem [{\citenamefont {Goodrich}\ \emph {et~al.}(2014)\citenamefont
  {Goodrich}, \citenamefont {Liu},\ and\ \citenamefont
  {Nagel}}]{goodrich2014solids}%
  \BibitemOpen
  \bibfield  {author} {\bibinfo {author} {\bibfnamefont {C.~P.}\ \bibnamefont
  {Goodrich}}, \bibinfo {author} {\bibfnamefont {A.~J.}\ \bibnamefont {Liu}},\
  and\ \bibinfo {author} {\bibfnamefont {S.~R.}\ \bibnamefont {Nagel}},\
  }\bibfield  {title} {\bibinfo {title} {Solids between the mechanical extremes
  of order and disorder},\ }\href {https://doi.org/10.1038/nphys3006}
  {\bibfield  {journal} {\bibinfo  {journal} {Nature Physics}\ }\textbf
  {\bibinfo {volume} {10}},\ \bibinfo {pages} {578} (\bibinfo {year}
  {2014})}\BibitemShut {NoStop}%
\bibitem [{\citenamefont {Kawasaki}\ and\ \citenamefont
  {Miyazaki}(2024)}]{kawasaki2024unified}%
  \BibitemOpen
  \bibfield  {author} {\bibinfo {author} {\bibfnamefont {T.}~\bibnamefont
  {Kawasaki}}\ and\ \bibinfo {author} {\bibfnamefont {K.}~\bibnamefont
  {Miyazaki}},\ }\bibfield  {title} {\bibinfo {title} {Unified understanding of
  nonlinear rheology near the jamming transition point},\ }\href
  {https://doi.org/10.1103/PhysRevLett.132.268201} {\bibfield  {journal}
  {\bibinfo  {journal} {Phys. Rev. Lett.}\ }\textbf {\bibinfo {volume} {132}},\
  \bibinfo {pages} {268201} (\bibinfo {year} {2024})}\BibitemShut {NoStop}%
\bibitem [{\citenamefont {Pan}\ \emph {et~al.}(2023)\citenamefont {Pan},
  \citenamefont {Wang}, \citenamefont {Yoshino}, \citenamefont {Zhang},\ and\
  \citenamefont {Jin}}]{pan2023review}%
  \BibitemOpen
  \bibfield  {author} {\bibinfo {author} {\bibfnamefont {D.}~\bibnamefont
  {Pan}}, \bibinfo {author} {\bibfnamefont {Y.}~\bibnamefont {Wang}}, \bibinfo
  {author} {\bibfnamefont {H.}~\bibnamefont {Yoshino}}, \bibinfo {author}
  {\bibfnamefont {J.}~\bibnamefont {Zhang}},\ and\ \bibinfo {author}
  {\bibfnamefont {Y.}~\bibnamefont {Jin}},\ }\bibfield  {title} {\bibinfo
  {title} {A review on shear jamming},\ }\href
  {https://doi.org/https://doi.org/10.1016/j.physrep.2023.10.002} {\bibfield
  {journal} {\bibinfo  {journal} {Physics Reports}\ }\textbf {\bibinfo {volume}
  {1038}},\ \bibinfo {pages} {1} (\bibinfo {year} {2023})}\BibitemShut
  {NoStop}%
\bibitem [{\citenamefont {Tong}\ \emph {et~al.}(2015)\citenamefont {Tong},
  \citenamefont {Tan},\ and\ \citenamefont {Xu}}]{tong2015crystals}%
  \BibitemOpen
  \bibfield  {author} {\bibinfo {author} {\bibfnamefont {H.}~\bibnamefont
  {Tong}}, \bibinfo {author} {\bibfnamefont {P.}~\bibnamefont {Tan}},\ and\
  \bibinfo {author} {\bibfnamefont {N.}~\bibnamefont {Xu}},\ }\bibfield
  {title} {\bibinfo {title} {From crystals to disordered crystals: A hidden
  order-disorder transition},\ }\href {https://doi.org/10.1038/srep15378}
  {\bibfield  {journal} {\bibinfo  {journal} {Scientific Reports}\ }\textbf
  {\bibinfo {volume} {5}},\ \bibinfo {pages} {15378} (\bibinfo {year}
  {2015})}\BibitemShut {NoStop}%
\bibitem [{\citenamefont {Ikeda}(2020)}]{ikeda2020jamming}%
  \BibitemOpen
  \bibfield  {author} {\bibinfo {author} {\bibfnamefont {H.}~\bibnamefont
  {Ikeda}},\ }\bibfield  {title} {\bibinfo {title} {Jamming and replica
  symmetry breaking of weakly disordered crystals},\ }\href
  {https://doi.org/10.1103/PhysRevResearch.2.033220} {\bibfield  {journal}
  {\bibinfo  {journal} {Phys. Rev. Res.}\ }\textbf {\bibinfo {volume} {2}},\
  \bibinfo {pages} {033220} (\bibinfo {year} {2020})}\BibitemShut {NoStop}%
\bibitem [{\citenamefont {Maher}\ \emph {et~al.}(2023)\citenamefont {Maher},
  \citenamefont {Jiao},\ and\ \citenamefont
  {Torquato}}]{maher2023hyperuniformity}%
  \BibitemOpen
  \bibfield  {author} {\bibinfo {author} {\bibfnamefont {C.~E.}\ \bibnamefont
  {Maher}}, \bibinfo {author} {\bibfnamefont {Y.}~\bibnamefont {Jiao}},\ and\
  \bibinfo {author} {\bibfnamefont {S.}~\bibnamefont {Torquato}},\ }\bibfield
  {title} {\bibinfo {title} {Hyperuniformity of maximally random jammed
  packings of hyperspheres across spatial dimensions},\ }\href
  {https://doi.org/10.1103/PhysRevE.108.064602} {\bibfield  {journal} {\bibinfo
   {journal} {Phys. Rev. E}\ }\textbf {\bibinfo {volume} {108}},\ \bibinfo
  {pages} {064602} (\bibinfo {year} {2023})}\BibitemShut {NoStop}%
\bibitem [{\citenamefont {Rissone}\ \emph {et~al.}(2021)\citenamefont
  {Rissone}, \citenamefont {Corwin},\ and\ \citenamefont
  {Parisi}}]{rissone2021long}%
  \BibitemOpen
  \bibfield  {author} {\bibinfo {author} {\bibfnamefont {P.}~\bibnamefont
  {Rissone}}, \bibinfo {author} {\bibfnamefont {E.~I.}\ \bibnamefont
  {Corwin}},\ and\ \bibinfo {author} {\bibfnamefont {G.}~\bibnamefont
  {Parisi}},\ }\bibfield  {title} {\bibinfo {title} {Long-range anomalous decay
  of the correlation in jammed packings},\ }\href
  {https://doi.org/10.1103/PhysRevLett.127.038001} {\bibfield  {journal}
  {\bibinfo  {journal} {Phys. Rev. Lett.}\ }\textbf {\bibinfo {volume} {127}},\
  \bibinfo {pages} {038001} (\bibinfo {year} {2021})}\BibitemShut {NoStop}%
\bibitem [{\citenamefont {Bitzek}\ \emph {et~al.}(2006)\citenamefont {Bitzek},
  \citenamefont {Koskinen}, \citenamefont {G\"ahler}, \citenamefont {Moseler},\
  and\ \citenamefont {Gumbsch}}]{bitzek2006structural}%
  \BibitemOpen
  \bibfield  {author} {\bibinfo {author} {\bibfnamefont {E.}~\bibnamefont
  {Bitzek}}, \bibinfo {author} {\bibfnamefont {P.}~\bibnamefont {Koskinen}},
  \bibinfo {author} {\bibfnamefont {F.}~\bibnamefont {G\"ahler}}, \bibinfo
  {author} {\bibfnamefont {M.}~\bibnamefont {Moseler}},\ and\ \bibinfo {author}
  {\bibfnamefont {P.}~\bibnamefont {Gumbsch}},\ }\bibfield  {title} {\bibinfo
  {title} {Structural relaxation made simple},\ }\href
  {https://doi.org/10.1103/PhysRevLett.97.170201} {\bibfield  {journal}
  {\bibinfo  {journal} {Phys. Rev. Lett.}\ }\textbf {\bibinfo {volume} {97}},\
  \bibinfo {pages} {170201} (\bibinfo {year} {2006})}\BibitemShut {NoStop}%
\bibitem [{\citenamefont {Majmudar}\ \emph {et~al.}(2007)\citenamefont
  {Majmudar}, \citenamefont {Sperl}, \citenamefont {Luding},\ and\
  \citenamefont {Behringer}}]{majmudar2007jamming}%
  \BibitemOpen
  \bibfield  {author} {\bibinfo {author} {\bibfnamefont {T.~S.}\ \bibnamefont
  {Majmudar}}, \bibinfo {author} {\bibfnamefont {M.}~\bibnamefont {Sperl}},
  \bibinfo {author} {\bibfnamefont {S.}~\bibnamefont {Luding}},\ and\ \bibinfo
  {author} {\bibfnamefont {R.~P.}\ \bibnamefont {Behringer}},\ }\bibfield
  {title} {\bibinfo {title} {Jamming transition in granular systems},\ }\href
  {https://doi.org/10.1103/PhysRevLett.98.058001} {\bibfield  {journal}
  {\bibinfo  {journal} {Phys. Rev. Lett.}\ }\textbf {\bibinfo {volume} {98}},\
  \bibinfo {pages} {058001} (\bibinfo {year} {2007})}\BibitemShut {NoStop}%
\bibitem [{\citenamefont {Tong}\ and\ \citenamefont
  {Tanaka}(2019)}]{tong2019structural}%
  \BibitemOpen
  \bibfield  {author} {\bibinfo {author} {\bibfnamefont {H.}~\bibnamefont
  {Tong}}\ and\ \bibinfo {author} {\bibfnamefont {H.}~\bibnamefont {Tanaka}},\
  }\bibfield  {title} {\bibinfo {title} {Structural order as a genuine control
  parameter of dynamics in simple glass formers},\ }\href
  {https://doi.org/10.1038/s41467-019-13606-3} {\bibfield  {journal} {\bibinfo
  {journal} {Nature Communications}\ }\textbf {\bibinfo {volume} {10}},\
  \bibinfo {pages} {5596} (\bibinfo {year} {2019})}\BibitemShut {NoStop}%
\bibitem [{\citenamefont {Goodrich}\ \emph {et~al.}(2012)\citenamefont
  {Goodrich}, \citenamefont {Liu},\ and\ \citenamefont
  {Nagel}}]{goodrich2012finite}%
  \BibitemOpen
  \bibfield  {author} {\bibinfo {author} {\bibfnamefont {C.~P.}\ \bibnamefont
  {Goodrich}}, \bibinfo {author} {\bibfnamefont {A.~J.}\ \bibnamefont {Liu}},\
  and\ \bibinfo {author} {\bibfnamefont {S.~R.}\ \bibnamefont {Nagel}},\
  }\bibfield  {title} {\bibinfo {title} {Finite-size scaling at the jamming
  transition},\ }\href {https://doi.org/10.1103/PhysRevLett.109.095704}
  {\bibfield  {journal} {\bibinfo  {journal} {Phys. Rev. Lett.}\ }\textbf
  {\bibinfo {volume} {109}},\ \bibinfo {pages} {095704} (\bibinfo {year}
  {2012})}\BibitemShut {NoStop}%
\bibitem [{\citenamefont {Binder}\ \emph {et~al.}(1985)\citenamefont {Binder},
  \citenamefont {Nauenberg}, \citenamefont {Privman},\ and\ \citenamefont
  {Young}}]{Binder1985prb}%
  \BibitemOpen
  \bibfield  {author} {\bibinfo {author} {\bibfnamefont {K.}~\bibnamefont
  {Binder}}, \bibinfo {author} {\bibfnamefont {M.}~\bibnamefont {Nauenberg}},
  \bibinfo {author} {\bibfnamefont {V.}~\bibnamefont {Privman}},\ and\ \bibinfo
  {author} {\bibfnamefont {A.~P.}\ \bibnamefont {Young}},\ }\bibfield  {title}
  {\bibinfo {title} {Finite-size tests of hyperscaling},\ }\href
  {https://doi.org/10.1103/PhysRevB.31.1498} {\bibfield  {journal} {\bibinfo
  {journal} {Phys. Rev. B}\ }\textbf {\bibinfo {volume} {31}},\ \bibinfo
  {pages} {1498} (\bibinfo {year} {1985})}\BibitemShut {NoStop}%
\bibitem [{\citenamefont {Wittmann}\ and\ \citenamefont
  {Young}(2014)}]{Wittmann2014pre}%
  \BibitemOpen
  \bibfield  {author} {\bibinfo {author} {\bibfnamefont {M.}~\bibnamefont
  {Wittmann}}\ and\ \bibinfo {author} {\bibfnamefont {A.~P.}\ \bibnamefont
  {Young}},\ }\bibfield  {title} {\bibinfo {title} {Finite-size scaling above
  the upper critical dimension},\ }\href
  {https://doi.org/10.1103/PhysRevE.90.062137} {\bibfield  {journal} {\bibinfo
  {journal} {Phys. Rev. E}\ }\textbf {\bibinfo {volume} {90}},\ \bibinfo
  {pages} {062137} (\bibinfo {year} {2014})}\BibitemShut {NoStop}%
\bibitem [{\citenamefont {Wang}\ \emph {et~al.}()\citenamefont {Wang},
  \citenamefont {Wang}, \citenamefont {Lei},\ and\ \citenamefont
  {Ma}}]{Wang2025d}%
  \BibitemOpen
  \bibfield  {author} {\bibinfo {author} {\bibfnamefont {H.-D.}\ \bibnamefont
  {Wang}}, \bibinfo {author} {\bibfnamefont {B.}~\bibnamefont {Wang}}, \bibinfo
  {author} {\bibfnamefont {Q.-L.}\ \bibnamefont {Lei}},\ and\ \bibinfo {author}
  {\bibfnamefont {Y.-Q.}\ \bibnamefont {Ma}},\ }\bibfield  {title} {\bibinfo
  {title} {Anomalous criticality of absorbing state transition toward
  jamming},\ }\href@noop {} {\ ,\ \bibinfo {pages}
  {arXiv:2510.06641}}\BibitemShut {NoStop}%
\bibitem [{\citenamefont {Wang}\ \emph {et~al.}(2025)\citenamefont {Wang},
  \citenamefont {Qian}, \citenamefont {Tong},\ and\ \citenamefont
  {Tanaka}}]{wang2025hyperuniform}%
  \BibitemOpen
  \bibfield  {author} {\bibinfo {author} {\bibfnamefont {Y.}~\bibnamefont
  {Wang}}, \bibinfo {author} {\bibfnamefont {Z.}~\bibnamefont {Qian}}, \bibinfo
  {author} {\bibfnamefont {H.}~\bibnamefont {Tong}},\ and\ \bibinfo {author}
  {\bibfnamefont {H.}~\bibnamefont {Tanaka}},\ }\bibfield  {title} {\bibinfo
  {title} {Hyperuniform disordered solids with crystal-like stability},\ }\href
  {https://doi.org/10.1038/s41467-025-56283-1} {\bibfield  {journal} {\bibinfo
  {journal} {Nature Communications}\ }\textbf {\bibinfo {volume} {16}},\
  \bibinfo {pages} {1398} (\bibinfo {year} {2025})}\BibitemShut {NoStop}%
\bibitem [{\citenamefont {Hexner}\ \emph {et~al.}(2018)\citenamefont {Hexner},
  \citenamefont {Liu},\ and\ \citenamefont {Nagel}}]{hexner2018two}%
  \BibitemOpen
  \bibfield  {author} {\bibinfo {author} {\bibfnamefont {D.}~\bibnamefont
  {Hexner}}, \bibinfo {author} {\bibfnamefont {A.~J.}\ \bibnamefont {Liu}},\
  and\ \bibinfo {author} {\bibfnamefont {S.~R.}\ \bibnamefont {Nagel}},\
  }\bibfield  {title} {\bibinfo {title} {Two diverging length scales in the
  structure of jammed packings},\ }\href
  {https://doi.org/10.1103/PhysRevLett.121.115501} {\bibfield  {journal}
  {\bibinfo  {journal} {Phys. Rev. Lett.}\ }\textbf {\bibinfo {volume} {121}},\
  \bibinfo {pages} {115501} (\bibinfo {year} {2018})}\BibitemShut {NoStop}%
\bibitem [{\citenamefont {Hexner}\ \emph {et~al.}(2019)\citenamefont {Hexner},
  \citenamefont {Urbani},\ and\ \citenamefont {Zamponi}}]{hexner2019can}%
  \BibitemOpen
  \bibfield  {author} {\bibinfo {author} {\bibfnamefont {D.}~\bibnamefont
  {Hexner}}, \bibinfo {author} {\bibfnamefont {P.}~\bibnamefont {Urbani}},\
  and\ \bibinfo {author} {\bibfnamefont {F.}~\bibnamefont {Zamponi}},\
  }\bibfield  {title} {\bibinfo {title} {Can a large packing be assembled from
  smaller ones?},\ }\href {https://doi.org/10.1103/PhysRevLett.123.068003}
  {\bibfield  {journal} {\bibinfo  {journal} {Phys. Rev. Lett.}\ }\textbf
  {\bibinfo {volume} {123}},\ \bibinfo {pages} {068003} (\bibinfo {year}
  {2019})}\BibitemShut {NoStop}%
  \bibitem [{\citenamefont {Maher}\ and\ \citenamefont {Torquato}(2024)}]{maher2024hyperuniformity}%
  \BibitemOpen
  \bibfield  {author} {\bibinfo {author} {\bibfnamefont {C.~E.}\ \bibnamefont
  		{Maher}}\ and\ \bibinfo {author} {\bibfnamefont {S.}\ \bibnamefont
  		{Torquato}},\ }\bibfield
  {title} {\bibinfo {title} {Hyperuniformity scaling of maximally random jammed
  		packings of two-dimensional binary disks},\ }\href
  {https://doi.org/10.1103/PhysRevE.110.064605} {\bibfield  {journal}
  	{\bibinfo  {journal} {Phys. Rev. E}\ }\textbf {\bibinfo {volume} {110}},\
  	\bibinfo {pages} {064605} (\bibinfo {year} {2024})}\BibitemShut {NoStop}%
\end{thebibliography}
%

\end{document}